\documentclass[%
reprint,
superscriptaddress,
amsmath,amssymb,
aps,
floatfix
]{revtex4-1}
\usepackage{graphicx}
\usepackage{color}
\usepackage{hyperref}
\usepackage{float}
\usepackage{array, multirow}
\usepackage{amsmath}
\usepackage{newtxtext}              
\usepackage{booktabs}
\usepackage{xcolor} 
\usepackage{soul}
\usepackage[caption=false]{subfig}
\usepackage[margin=0.8in]{geometry} 
\usepackage{todonotes}
\makeatletter \renewcommand\@make@capt@title[2]{%
\@ifx@empty\float@link{\@firstofone}{\expandafter\href\expandafter{\float@link}}%
\sffamily{\textbf{#1}}\@caption@fignum@sep#2 }

\hypersetup{colorlinks, 
    linkcolor={blue!75!black!80!yellow},
    citecolor={blue!75!black!80!yellow}, 
    urlcolor={blue!75!black!80!yellow}
    }
\begin{document}

\title{Purification and Entanglement Routing on Quantum Networks}

\author{Michelle Victora}
\email{michelle@aliroquantum.com}
\affiliation{Aliro Technologies, Inc.}
\author{Stefan Krastanov}
\affiliation{John A. Paulson School of Engineering and Applied Sciences, Harvard
University, Cambridge, MA 02138, USA}
\affiliation{Department of Electrical Engineering and Computer Science, Massachusetts Institute of Technology, Cambridge, MA 02139, USA}
\author{Alexander Sanchez de la Cerda}
\affiliation{John A. Paulson School of Engineering and Applied Sciences, Harvard
University, Cambridge, MA 02138, USA}
\author{Steven Willis}
\affiliation{Aliro Technologies, Inc.}
\author{Prineha Narang}
\email{prineha@seas.harvard.edu}
\affiliation{John A. Paulson School of Engineering and Applied Sciences, Harvard
University, Cambridge, MA 02138, USA}

\date{\today}

\begin{abstract}
\noindent We present an approach to purification and entanglement routing on complex quantum network architectures, that is, how a quantum network equipped with imperfect channel fidelities and limited memory storage time can distribute entanglement between users. We explore how network parameters influence the performance of path-finding algorithms necessary for optimizing routing and, in particular, we explore the interplay between the bandwidth of a quantum channels and the choice of purification protocol. Finally, we demonstrate multi-path routing on various network topologies with resource constraints, in an effort to inform future design choices for quantum network configurations. Our work optimizes both the choice of path over the quantum network and the choice of purification schemes used between nodes. We consider not only pair-production rate, but optimize over the fidelity of the delivered entangled state. We introduce effective heuristics enabling fast path-finding algorithms for maximizing entanglement shared between two nodes on a quantum network, with performance comparable to that of a computationally-expensive `brute-force' path search. 
\end{abstract}

\maketitle
\section{Quantum Network}
\label{sec:quantumnetwork}

\subsection{Introduction}
\noindent 

A quantum network \cite{vanmeternetworking,kimbleinternet,castelvecchiinternet,wehnerinternet} is used to generate, distribute, and process quantum information for a variety of applications. For many use cases, the most important prerequisite is the generation of long-distance entanglement between users, whether that be for distributed quantum computing \cite{beyondqkd, distributedqc2, distributedqc3}, multiparty cryptography \cite{multipartysecrets}, improved sensing \cite{sensing1, sensing2}, and blind quantum computation \cite{blindqc}. Already, recent experiments have demonstrated successful long-range entanglement links, from terrestrial point-to-point links over a distance of 1000 km \cite{longrangeentanglement}, to satellite-ground entanglement distribution \cite{satelliteentanglement1,satelliteentanglement2}. As the field has defined in recent `blueprints' \cite{quantuminternetishere}, creating repeating, switching, and routing technologies for quantum entanglement is a central pillar towards the development of a quantum network. Furthermore, performing error detection on \emph{quantum repeaters} \cite{repeaters2} using entanglement purification \cite{repeaters1, PhysRevA.59.169} is another central pillar of network development which has a significant impact on routing protocols, as we will demonstrate in this paper. Here, we explore how error detection and routing can inform each other as we aim to design a high-functioning quantum internet with real-world network parameters.

 Quantum networks can be characterized by three main components -- the end nodes (quantum processors that can receive and emit information), channels (classical and quantum communication lines over long distances), and repeaters (intermediate nodes along a communication line that act to manage decoherence across the network). By combining these components strategically, quantum networks can link multiple quantum computers to make a larger Hilbert space. Quantum networking can also be used to enable highly secure quantum communication using quantum cryptographic protocols. Currently, there is a large research effort to develop a quantum network stack \cite{quantumnetworkstack} for reliably running protocols on entanglement-based networks. We contribute to this development by demonstrating the performance of heuristic link costs for routing protocols on a quantum network with imperfect processor fidelities and limited memory coherence time using optimized purification protocols.

Purification protocols were discussed as early as 1996 by Charles Bennett et al. in \cite{bennett1996concentrating}, which introduced the idea of using local operations on $n$ entangled pairs to increase the fidelity of a single pair. A quantum repeater chain uses a two-qubit Bell-State measurement (BSM) that is used to perform  \emph{entanglement swapping} \cite{BSM1, BSM2, BSM3}. which essentially glue together small entanglement links into longer-distance entangled links between multiple parties \cite{saikatrouting}. Briegel and D{\"u}r developed methods of performing \emph{purification} on a quantum repeater chain \cite{repeaters1, PhysRevA.59.169} in a nested fashion to detect errors introduced in the connection process.   Entanglement pumping \cite{PhysRevA.59.169} and Bennett's recurrence protocol \cite{PhysRevLett.77.2818} provide a method for converting $m$ pairs into a single pair of asymptotically high fidelity, limited only by the gate fidelity of the operators \cite{doubleselection}. Optimizing the use of these protocols across repeater chains results in a non-trivial problem as resource constraints, channel length, and input state quality all affect the purification strategies \cite{PhysRevA.87.062335}. While entanglement pumping and recurrence protocols allow one to achieve an asymptotically high state fidelity \cite{PhysRevLett.76.722}, repeated gate operations can result in marginal returns in the latter stages of purification. To improve upon these schemes, much work has gone into developing and optimizing new purification schemes using $m>2$ pairs, applying error correcting codes to explore a wider range of circuit options\cite{aschauer2005}. Recently, work by Krastanov et al. \cite{stefanpurification} developed a method for optimizing purification circuits with respect to circuit width, gate, and input state fidelity, utilizing permutation schemes \cite{PhysRevA.67.022310, PhysRevA.72.032313} to increase the conversion efficiency of conversion.

In addition to purification at the local level, much recent work has analyzed the complicated problem of entanglement routing across a quantum network, with the goal of maximizing quality entanglement generation between source and destination. Schoute et al. \cite{schouteroutingtopologies} developed routing protocols on specific network topologies and found scaling laws under the assumption that each link generates a perfect, lossless EPR pair in  every time slot and each repeater's actions are limited to perfect Bell state measurements (BSM's). Guha et al. \cite{saikatrouting} developed local-knowledge and global-knowledge quantum network routing protocols using a simplified model where the only source of imperfection is pure loss. Wehner et al. \cite{chakrabortydistributedrouting} used a similar pure-loss model to test performance of adapted classical networking routing algorithms on a quantum network under the strain of multiple concurrent requests. Wehner et al.   constructed an efficient linear-programming formulation, where they approached entanglement routing using a multi-commodity flow-based approach with perfect gate operations and  imperfect channels without purification \cite{wehnermulticommodity}. Van Meter et  al. \cite{vanmeterrouting} analyzed topologically complex networks and defined metrics for quantifying total work along a path and informing path selection. In this paper, we optimize routing and purification protocols for distributing entanglement among source-destination pairs. The novel approach of this \emph{Article} is to combine both routing and custom purification protocols in a quantum internet with an emphasis on noisy gates, lossy channels between quantum repeaters, and finite memory storage time.

For each section of our \emph{Article}, we evaluate our results with both perfect and imperfect gate fidelities in order to highlight the effect of gate imperfections across a network, which has not been studied in the context of routing but is fundamental to the analysis of quantum networks. In Section \ref{sec:puropt}, we review the concept of purification and introduce our optimized purification routine for a finite-memory repeater chain made up of channels with varied entanglement generation rates. We then discuss our routing work in Section \ref{sec:routing}, in which we optimize path selection on a quantum network. We conclude with Section \ref{sec:topk}, where we compare the performance of a greedy multi-path routing scheme on quantum networks with varying topologies.

\begin{figure*}[!tbp]
    \includegraphics[width=.7\textwidth]{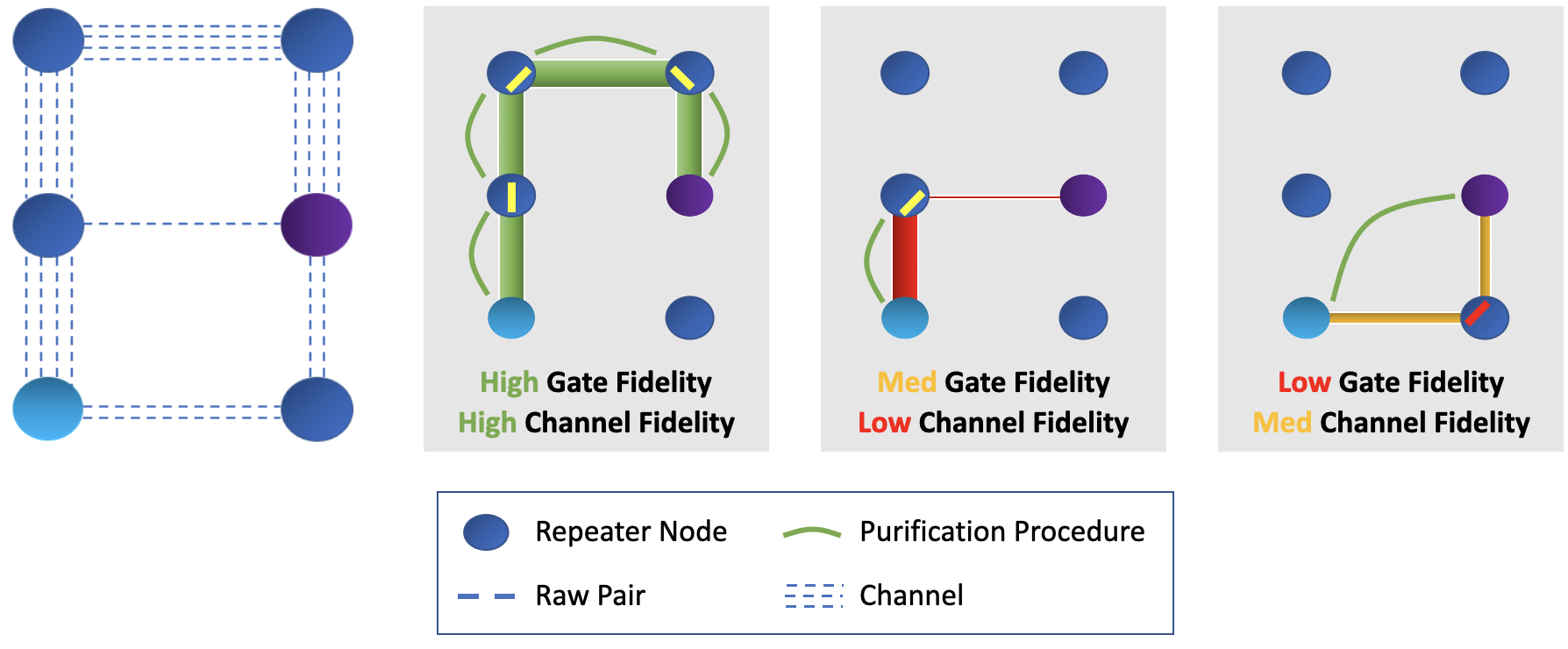}
    \label{fig:f2}
  \caption{\textbf{(Left-to-right)} The network over which we search for the best entanglement distribution procedure; Under high gate and channel fidelities, routing can opt for longer paths with higher throughput; When limited by channel fidelity, shorter paths become more attractive, with high-bandwidth channels between nodes allowing more selective purification; With gate fidelity as a limiting factor, purification is best for correcting across multiple hops, leading to a need for near-identical neighboring channels.}
  \label{fig:pathcapacity}
\end{figure*}


\subsection{Model}
\label{subsec:networkmodel}

To model a quantum network, we can consider a 2D graph \emph{G(V, E)} as the topology for our repeater network. Each node $v \subseteq V$ is a repeater, with each edge $e \subseteq E$ representing a physical link acting as the communication line, or \emph{channel}, between two repeater nodes. The network is synchronized to a clock where each timestep is no longer than the memory decoherence time. $E$ is characterized by its entanglement generation rate, i.e., EGR. Specifically, for each timestep, EGR$_{u, v}$ is the number of entangled pairs generated between two nodes $u, v \subseteq V$ where $u$ and $v$ each share a half of the entangled state. The EGR will necessarily be a function of source brightness, fiber loss, coupling efficiency, and qubit buffer size of the network node. We treat this generation as deterministic for this study, as we are specifically interested in how these varied rates affect expected distillable entanglement across a particular path. Physically, \emph{near-deterministic} elementary pair generation can be achieved using temporal and spectral multiplexing on linear-optics and solid-state/atomic-ensemble memories \cite{multiplex}. 

  Quantum repeaters rely on quantum memories for storage of entangled pairs while purification and swapping processes are implemented. All necessary operations must be accomplished across all repeaters used within the duration of the time in which the memory is coherent and $t_{\mathrm{decoh}}$ is the time at which the memory decoheres. We model this as a discrete timestep, meaning that, for $t<t_{\mathrm{decoh}}$, the memory holds the received qubit with a static fidelity, and for $t>t_{\mathrm{decoh}}$, the qubit is lost. All necessary operations across the network must be completed before $t_{\mathrm{decoh}}$. In this paper we do not perform more than one round of purification among the many rounds of entanglement swapping. This scheme was explored greatly by Bruß et al. in \cite{PhysRevA.87.062335}, where it was shown that for high gate fidelities and low channel fidelities, this method produced higher distillable entanglement per memory per second than a standard nested purification scheme for an arbitrary repeater chain. We extend this scheme by allowing a single round of purification to extend across multiple hops. In all cases, the round-trip communication time involved in communicating these gate operations must be less than the decoherence time of our memories. We assume that all memories are on-demand memories, i.e., they can retrieve and release the stored entangled pairs whenever required. All repeaters have the capability of performing two-qubit and one-qubit gate operations, conditioned on measurement outcomes. 

To optimize routing across a near-term quantum network, we need to consider the accumulated decoherence across imperfect channels and repeaters. If noise across the network, which is introduced by imperfect channels and memories, leads to imperfect raw Bell pairs, then each \emph{entanglement swap} operation \cite{BSM1, BSM2, BSM3} that concatenates single-hop segments will result in a single segment of a lower fidelity than either of its two components. For the purposes of this work, we characterize any noise as complete \emph{depolarization}, as this forms the worst-case assumption of noise across a network. A pure state $|\psi_{00}\rangle$ that mixes with isotropic noise can be written as the canonical \emph{Werner state} \cite{werner}, written as:

\begin{equation}
  \rho = W|\psi_{00}\rangle\langle\psi_{00}| + \frac{(1-W)}{4}\textbf{I}_4
\end{equation}

where $|\psi_{00}\rangle = \frac{1}{\sqrt{2}} (|00\rangle + |11\rangle$ and the fidelity of our state to $|\psi\rangle$ is $F=\frac{3W+1}{4}$. These states are produced in intermediate entanglement source nodes between quantum repeaters, with each half of the Werner state sent to the respective adjacent quantum repeater. Each half is subsequently connected to another Werner state through a swapping procedure, either pre- or post-purification, resulting in a longer pair with an entanglement fidelity reflective of the chosen operations.  Entanglement swapping occurs on repeaters with unit probability, but with fidelity limited to gate operations, the likes of which can be achieved on the same kind of superconducting circuits on which we would be able to run our purification protocols \cite{biao2019swap}.  Swapping is a 1:1 relation, meaning that, for $n, m$ pairs available for connection through a swapping procedure, only $\min(n, m)$ new pairs will be generated. Two-way communication is used to communicate successful swapping.

The output states of these operations are repeatedly brought back to a Werner state with identical fidelity using a series of random bilateral rotations \cite{dur1999quantum}. If we consider simply a chain of $n$ neighboring Werner states that are connected in this method, we find that the fidelity of the generated state is:
\begin{align}
 F &= \frac{1}{4} + \frac{3}{4}\left(\frac{p_2(4\eta^2-1)}{3}\right)^{n-1}\times \nonumber\\
 &\left(\frac{4F_1-1}{3}\right)\left(\frac{4F_2-1}{3}\right)...\left(\frac{4F_n-1}{3}\right) \label{eq:degradation}
\end{align}
where $p_2$ is the two-qubit gate fidelity, $\eta$ is the measurement fidelity of the swapping operation, and $F_i$ represents the fidelity of the entanglement across a particular link. In this case $p_2$ is defined as the probability of depolarization occurring during the gate operation, and $\eta$ is the probability of a measurement reporting the incorrect result. In this paper, we compare perfect gate fidelities ($p_2= \eta=1$) and imperfect gate fidelities ($p_2 = \eta = 0.99$). No single-qubit errors are treated in this model as they are generally much lower.
 
 In the rest of the \emph{Article}, we will explore optimized routing and purification in a quantum network with varying entanglement generation rates between repeaters. Figure \ref{fig:pathcapacity} visualizes such a network as channels connecting neighboring repeaters vary in the number of pairs generated in a single timestep of the network. We define three different network parameter regimes where different routing strategies perform better. In near-perfect networks, we find that imperfect gate and channel fidelities may be small enough to enable a search for high-EGR (but potentially longer) paths. Low channel fidelity leads to accumulated decoherence that constrains the overall path length. By using more selective purification circuits on high-EGR channels, the network can increase the fidelity of individual channels in an effort to reduce decoherence across the network. For a network primarily limited by gate fidelity, purification is more advantageously performed post-BSM, and shorter paths with neighboring channels of near-identical entanglement generation rates will better sustain purification across these connected channels.

\section{Purification Optimization}
\label{sec:puropt}

\subsection{Introduction to Purification}
\label{subsec:purint}

In order to demonstrate our routing work in Section \ref{sec:routing}, we must first optimize our purification routines to a quantum repeater chain  with varied entanglement generation rates across channels. Purification protocols consist of performing local operations on \emph{n} entangled pairs shared between two parties, resulting in a smaller number of higher-fidelity pairs. Typically, local operations combined with CNOT gates are applied, with different sequences depending on the specific protocol. Once all gate operations are complete, both parties perform a measurement and, depending on the outcome, the resulting pair is determined to have a higher fidelity or it is discarded. These protocols are probabilistic, leading to reduced rates from the initial entanglement generation rate of raw pairs. The resulting fidelity gain from a circuit is increased with the number of pairs sacrificed. From here on out, we will define circuits that sacrifice a greater fraction of raw pairs as more \emph{selective} circuits.  We will run these circuits in parallel on $n$ pairs so as to generate $m$ entangled pairs where $m<n$.

In this \emph{Article}, we assume quantum repeaters that have the resources and are capable of performing both purification and entanglement swapping \cite{muralidharan2016optimal, wehnerinternet}. Specifically, we consider quantum protocols that perform only one purification round among its multiple entanglement swapping rounds. By allowing for flexibility in the stage at which purification occurs, we can perform the error detection at the point of largest impact.

Such purification schemes are a fundamental operation for quantum repeaters with limited memory coherence times; as the quantum internet develops, more advanced quantum repeaters using nested purification can be developed using the strategies presented in this paper and will complement the former. We leave the study of fidelity-optimized nested purification and routing schemes building on the resource-efficient approach developed in this paper for future work.

\subsection{Entanglement Generation Rate (EGR)}
\label{subsec:purmod}

We will be considering routing across a network in the following section, but in this section we use simple quantum repeater chains chains to introduce a number of necessary concepts.  We need to consider two constraints of our network: a) the discrete memory decoherence time, and b) differing entanglement generation rates (EGRs) across channels. The memory decoherence time limits our ability to perform time-intensive nested purification protocols, while the varied EGR means that purification must adapt to accommodate the difference across channels. Suppose we have a repeater chain of $(s, u_1, u_2, ... u_n, d)$, where all repeaters $u_i, u_{i+1}$ are neighbors of each other (i.e., connected by a channel) and $s$ would like to share entangled pairs with $d$. First, all repeaters generate entanglement with their neighbors/neighbor in parallel and store the entangled links in memory. An intermediate repeater, $u_i$ will perform either a swap or a purify-and-swap operation when repeaters $u_{i-1}$ and $u_{i+1}$ are ready. The overall entanglement generation rate with this protocol is lower than the raw EGR of one link. Assuming a chain with $n$ repeaters between  source $s$ and destination $d$ where each channel between repeaters $u$ and $v$ has a \emph{raw} entanglement generation rate EGR$(u, v)$, the expected entanglement generation rate $r_e$ will be such that:

\begin{equation}
    r_e \leq \min\{\text{Pur}(s, u_1),...\text{Pur}(u_{n},d)\}.
\label{eq:entanglementrate}
\end{equation}

\begin{figure}
\centering
  \centering
  \includegraphics[width=1\linewidth]{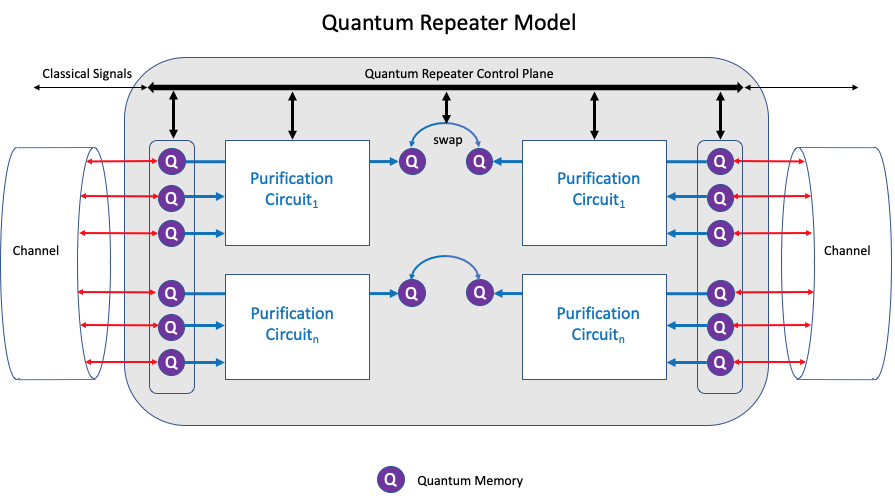}
  \caption{A diagram representing the structure of a quantum repeater from our model. Stored qubits from adjacent channels are operated on by purification circuits, reducing the overall number of purified qubits to be swapped for generating long-range entanglement.}
  \label{fig:repeater}
\end{figure}

Here, Pur$(s, u)$ represents the raw EGR post-purification. Specifically, $Pur \leq p_{succ}\times\left\lfloor{\text{EGR}/k}\right\rfloor$, where $k$ denotes the number of pairs to be used in a particular circuit, and $p_{succ}$ denotes the probability of success for that purification circuit to return a state with a higher fidelity. If a simple swapping operation is performed with no purification, then Pur = EGR. By decreasing the selectivity of circuits on lower-EGR channels and increasing the selectivity of circuits on higher-EGR channels, we can strike a balance of optimized EGR while still maintaining a usable fidelity. In this case, we do not include the time of circuit operation in our rate calculation as we assume all operations are able to take place in one timestep of our network. A repeater structure that implements this operation is shown in Figure \ref{fig:repeater}, with multiple qubit ``blocks" being designated to a single purification circuit of a specified $k$. In the following section, we discuss how we the balance between fidelity and EGR necessary for maximizing overall information flow.

\subsection{Distillable Entanglement}
\label{subsec:quantband}

A primary goal in quantum networking is to ensure the maximum information flow from source to destination. In quantum communication, the number of entangled pairs is not a complete metric to describe the information flow in a quantum network. Decoherence of each pair will reduce the quality of entanglement, which must be mitigated through purification. In this subsection, we discuss the maximization of quantum information flow between the source and a destination over a quantum repeater chain. Section \ref{sec:routing} will discuss how we consider routing in a quantum network as a repeater chain selection problem.

In order to correctly describe the information flow with respect to both the pair number and decoherence, we use a metric of distillable entanglement, $D(W)$, which represents the amount of purified pairs (near-perfect states) the source and destination can distill from $m$ shared impure pairs using local unitary operations and one-way communication. The functional form, derived by Bennett et al. is \cite{hashingrate}
\begin{equation}
    D(W) = m(1 + F\log_2 F + ((1-F)/3)\log_2(1-F)
\label{eq:distillent}
\end{equation} 
 where $m$ is the number of entangled pairs with a certain fidelity $F$.  We use this distillable entanglement as a metric for optimization. This way, we have the metric that accounts for both entanglement generation rate and the fidelity of our state into one functional form that allows us to effectively compare different repeater chains and purification schemes. $D(W)$ is a metric that is used in existing applications, such as the distillable key rate in a six-state quantum key distribution protocol \cite{wehner2020keyrates}.
 
 While both fidelity and entangled pair rate play a role in determining the overall distillable entanglement, these two parameters have an inverse relationship, as in order to increase the fidelity of one entangled pair, we must reduce the overall distillable pair number shared through purification. Quantum network protocols must focus on maintaining high-fidelity entanglement while not excessively reducing the EGR through overly selective purification schemes. The next section will describe how we balance these two competing parameters over a repeater chain of varying EGR.

 \subsection{Circuit Optimization}
 \label{subsec:purmethod}
 
 We model our decoherence across a chain  of repeaters using equation \ref{eq:degradation} combined with purification performed at various stages. This chain of repeaters is generated using an EGR uniformly spread from 8 to 32 raw entangled pairs per channel per timestep. We first build a family of circuits that span different node sizes and are optimized for varying gate fidelities. We then perform an exhaustive search to determine a distillation and entanglement swap protocol that provides the highest distillable entanglement on the generated chain. 
 
     \begin{figure}
\centering
  \centering
  \centering
  \includegraphics[width=1\linewidth]{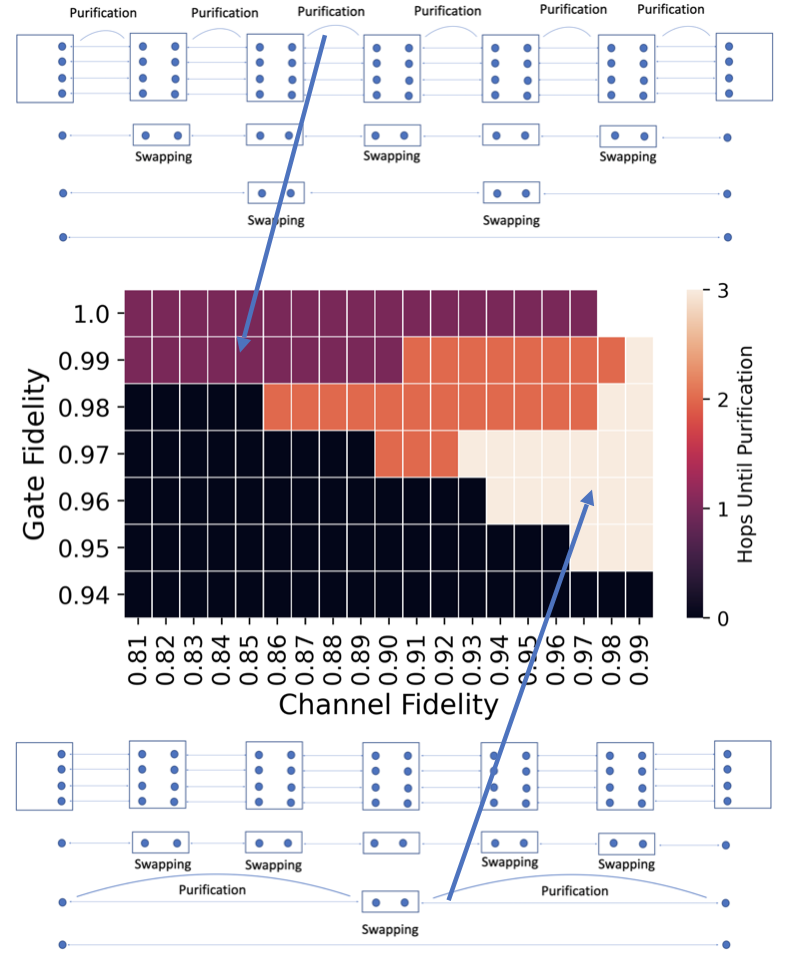}
    \centering
  \centering
  \caption{Colormap coded by number of hops over which optimal purification is performed for a path length of six hops for varying gate and channel fidelity. Uniform channel EGR of 20 pairs/timestep. 0's in the bottom left refer to no distillable entanglement achieved for any purification scheme explored. It should be noted that lower gate fidelity led to purification being optimally performed after a greater number of hops. This is a result of the constraint of purification only being performed for a single round. More swaps will lead to decreasing gate fidelity -- optimal entanglement rates will be achieved from purification correcting for as many swaps as possible. Missing boxes mean no purification was needed. Diagrams above and below represent what the full swapping and purification scheme will be for the different numbers of hops.}
  \label{fig:nestspread}
\end{figure}

For our purification optimization procedure, we first assign all channels the most selective distillation circuit we consider in this work, where $k=8$ pairs are converted into one single high-fidelity pair. We chose $k=8$ as our maximum selectivity, as increasingly selective circuits did not provide significantly increased distillable entanglement for our explored network parameters and repeater chain lengths.  We then perform a ``purification relaxation" process where we decrease the circuit selectivity across the channel/s with the lowest pair number post-purification, Pur$(u, v)$. Without relaxation, these channels essentially act as bottlenecks on the overall expected entanglement generation rate $r_e$ (see Equation \ref{eq:entanglementrate}), as $r_e$ is limited by the minimum channel EGR post-purification. By relaxing circuit selectivity across these channels, we increase the overall $r_e$ of the chain, but at a reduced final fidelity. This relaxation process is continued until we determine an an upper bound on achievable distillable entanglement (see Equation \ref{eq:distillent}). 

 In this work, we develop a library of purification circuits optimized with respect to gate and measurement fidelity using \href{https://github.com/Krastanov/qevo}{optimization software} written by Krastanov et al. \cite{stefanpurification}. Our channel EGR parameter includes consideration of a memory buffer large enough to hold all entangled pairs generated in a single timestep, thereby allowing us to consider the advantage of wide circuits used in near-perfect gate conditions. While many works have studied purification using standard protocols such as DEJMPS and BBPSSW \cite{dur2007entanglement}, we use circuits that are optimized to our network parameters so we can truly evaluate distillable entanglement in the limit of circuit gate and channel fidelity. 
 
 We consider both the purification circuits to use and the timing of purification (pre- or post-BSM) in this work. Additionally, if purification is best performed post-BSM, we must determine the number of hops over which to perform the purification protocol. We need to time our resource-intensive purification protocol so as to achieve the maximum fidelity increase with minimal qubits sacrificed. As stated in Section \ref{subsec:networkmodel}, we limit the distance across which purification can be performed to three hops in order to ensure all communication time fits within the timestep of the network. Each permutation of potential purification channel lengths is fed into our optimization scheme, along with our pre-BSM chain. In this way, our optimization protocol returns information both on circuit type and circuit timing.

 \subsection{Optimized Fidelity Results}
 \label{subsec:purres}
 
 Figure \ref{fig:nestspread} shows the number of hops after which purification was performed for a uniform  quantum repeater chain, where channels between nodes each produced 20 raw Bell pairs per timestep for a repeater separation of six hops. We observe the number of hops increase as gate fidelity drops, representing the preference for purification to correct for accumulated errors, in order to generate the most drastic improvement in fidelity at minimum number of sacrificial pairs. Additional plots detailing the distillable entanglement achieved for various entanglement generation rates are included in our Supplementary Material.
 
  As we introduce non-uniform quantum repeater chains, where channels between nodes have differing entanglement generation rates, we use asymmetric purification protocols that outperform symmetric protocols on a repeater chain with edges of varied entanglement generation rates. Specifically, we allow for different purification circuits to act across different edges of a selected repeater chain rather than enforcing a uniform purification routine. Considering optimized purification routines for repeater chains with varied entanglement generation rates is important as we consider routing in a quantum network where such repeater chains materialize as paths connecting source to destination. 
 
 Finally, it should be noted that we have simplified our problem in a few different ways. First, we have assumed deterministic entanglement generation between repeaters rather than a probabilistic distribution. Second, our decoherence model is a static fidelity for a discrete timestep, rather than an exponential decay. Finally,  we only consider isotropic noise in this model, rather than biased noise that we could target with more granularity in the goal of improving purification schemes. Even with these abstractions, however, we find that purification optimization across a repeater chain is non-trivial, and we aim here to provide a framework for optimizing information flow across a network using more comprehensive models.

\section{Entanglement Routing}
\label{sec:routing}

\subsection{Motivation}
\label{subsec:routingmot}

The previous section discussed optimizing purification protocols across repeater chains with edges of varying EGR. Quantum networks essentially contain a number of potential repeater chains to connect source to destination, and choosing the repeater chain can be just as important as choosing the correct purification procedure. We can think of routing as repeater chain selection, given the available chains that connect source to destination. This becomes a path selection problem. 


\subsection{Problem Statement}
\label{subsec:routingprob}

Since distillable entanglement is a function of both EGR and fidelity, path selection cannot be reduced to a simple shortest-distance or max-flow problem. As visualized in Figure \ref{fig:pathcapacity}, different network parameter regimes may require different routing methods that maximize the distillable entanglement shared between two repeaters. However, to compare all possible paths across a network while customizing a purification routine for each one becomes an exponentially difficult problem.  By developing an appropriate cost function to select for paths that have a greater overall EGR, we open up the possibility of using Dijkstra's weighted algorithm to reduce the search time to find a decent path.  It is not at all obvious whether a weighted path algorithm has a place in optimizing such a complex problem, particularly since that approach requires the overall cost function used to be a summation of individual edge costs, which does not match the functional form of distillable entanglement (Eq. \ref{eq:distillent}). 

As a note, a classical networking analogy to this problem would be the issue of allocating capacity with path length constraints on a network \cite{zhang2019constrainedrouting}. This tackles the issue of maximizing overall flow in a network when a path length constraint is imposed due to delay, jitter, or signal quality constraints. In our model, path length constraints are imposed by entanglement decoherence while these constraints are loosened by performing purification schemes that can reduce this decoherence. While here we talk about finding a single optimal path, we discuss in Section \ref{sec:multiroute} how we can implement multi-path routing to increase overall entanglement rate on a network over that which can be achieved on a single repeater chain.

\subsection{Path Selection}
\label{subsec:routingmethod}

\begin{figure}
\centering
  \centering
  \includegraphics[width=1\linewidth]{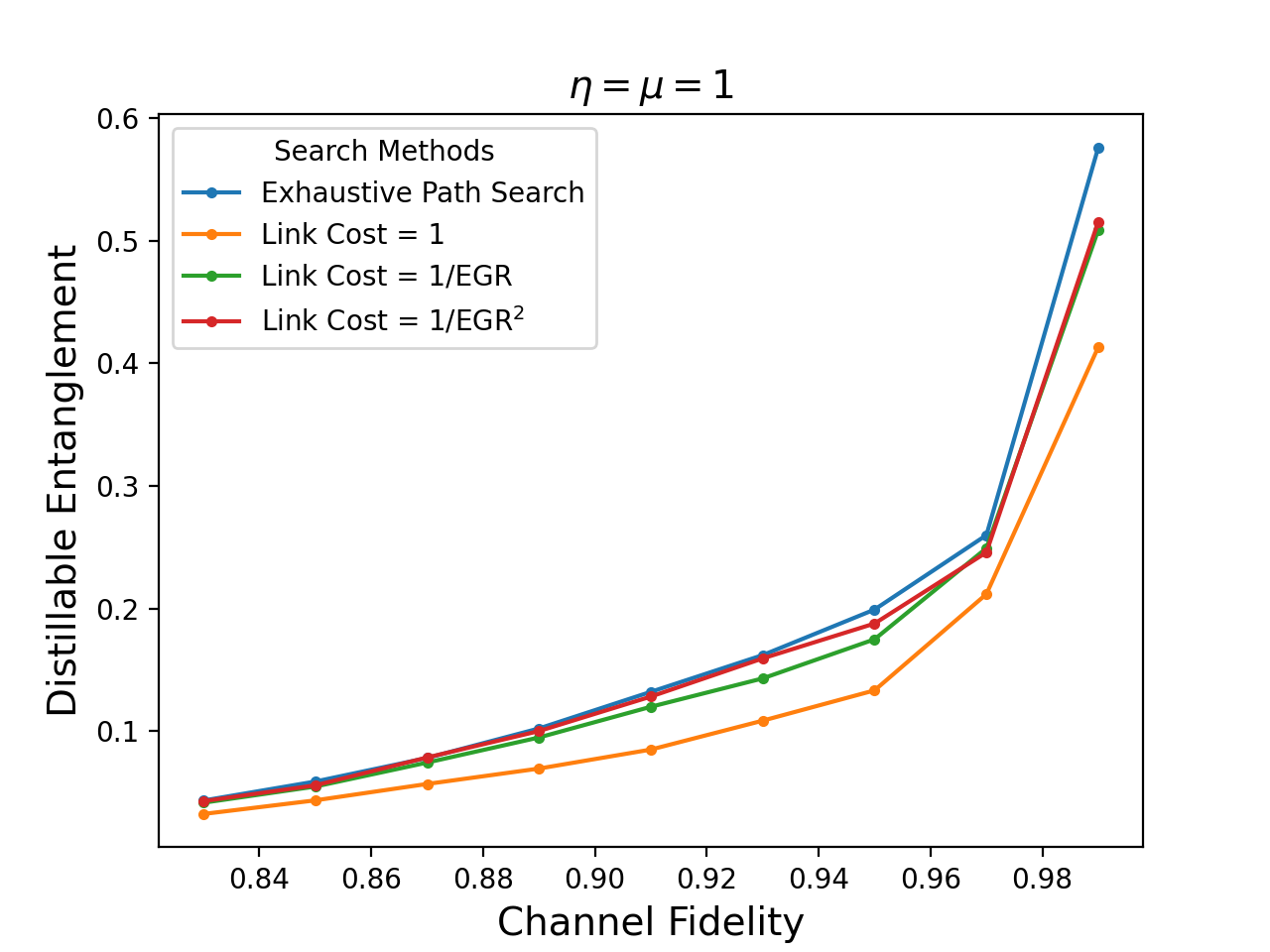}
  \centering
  \includegraphics[width=1\linewidth]{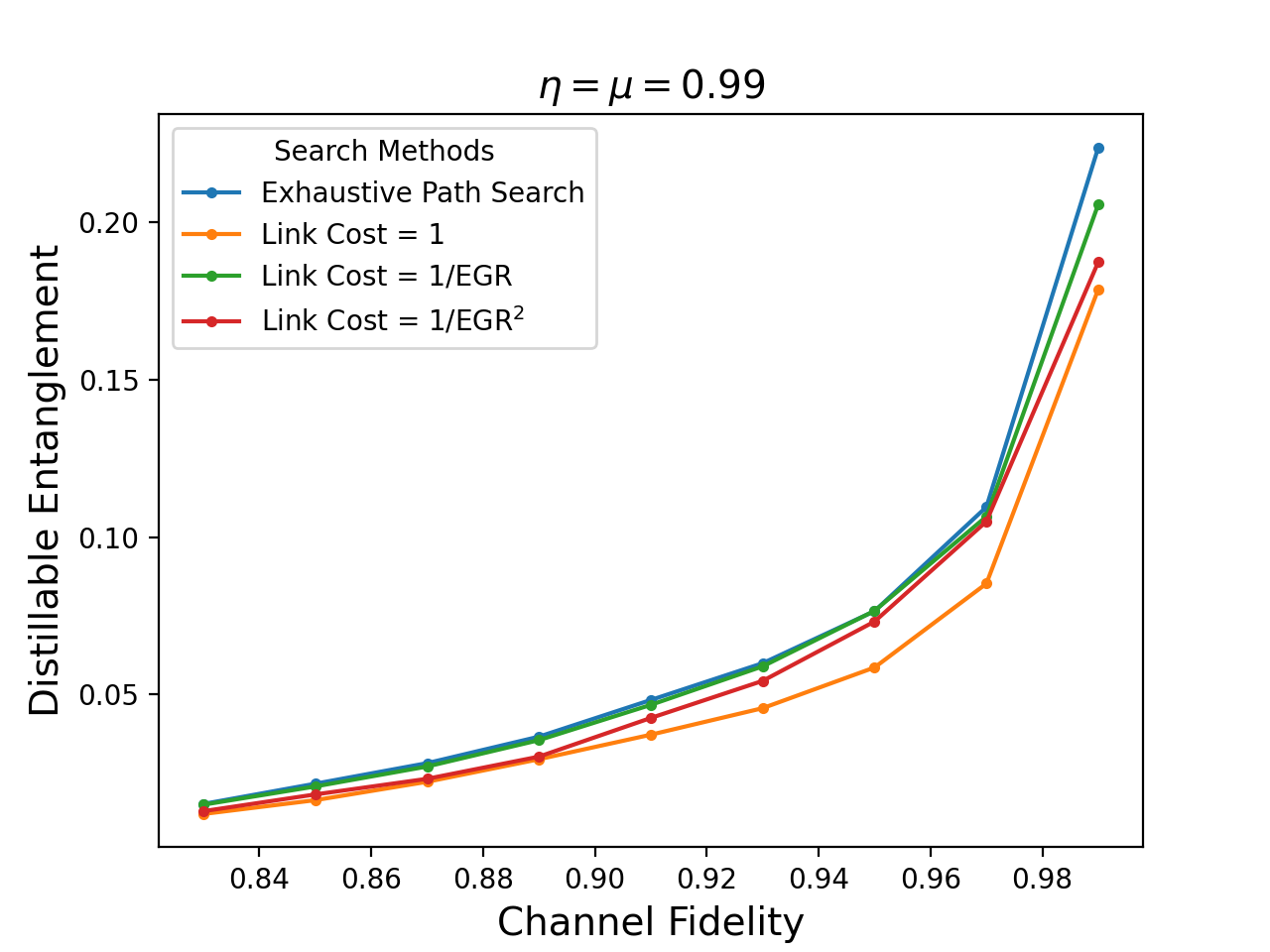}
  \caption{A comparison of distillable entanglement (normalized by average channel EGR) achieved using an exhaustive path search vs. Dijkstra's weighted algorithm on a triangular grid over a repeater separation of four hops. EGR spread from 8-32 raw entangled pairs per channel per timestep. Gate and measurement fidelity = 1 (top) and 0.99 (bottom). Similar comparisons for path searches on a square grid network and a hexagonal grid network are displayed in the supplementary material. Note the change in y-axes between graphs. }
  \label{fig:costcomparison}
\end{figure}
\begin{figure}
\centering
  \centering
  \includegraphics[width=0.7\linewidth]{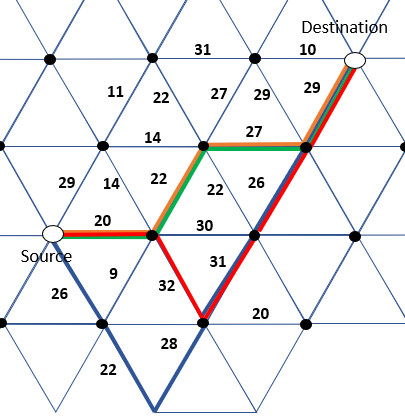}
  \caption{An example of four different paths found on a triangular grid, with path colors corresponding to paths found by methods listed in Figure \ref{fig:costcomparison}. Network parameters for optimal path search found was gate fidelity = measurement fidelity = 1, and channel fidelity = 0.99. Numbers listed by edges detail channel EGR. Path found by exhaustive path search implies a search for maximal minimum EGR along a path, while still balancing path length constraints.}
  \label{fig:trianglepaths}
\end{figure}

To gain insight on constructive useful heuristics for path selection, we first performed a search over all possible paths on a network with a cutoff length of 10 hops. The cutoff length was selected to reduce computational time. The longest repeater chain length we explore in this work is eight hops. By allowing an exhaustive search, we can explore how varying parameters of path length and path structure interplay with varied purification schemes that may span one to many hops. 

\subsection{Heuristic Path Selection}

With our method from the previous section, we can find a path with maximal distillable entanglement between source to destination on a network. By using a Dijkstra's weighted algorithm with an appropriate link cost, we can converge to a solution without needing to examine all possible paths. To determine an appropriate cost function for our channels, we compared three different channel costs. We used a link cost of $1$ to signify a search for the shortest path from source to destination. We compared this to a link cost of 1/EGR to signify \emph{inverse EGR}, and a link cost of 1/EGR$^2$ to signify \emph{inverse EGR squared}. Both of these latter functions will weight (with different degrees) towards channels that have much higher entanglement generation rates. While entanglement generation rate certainly has a logical place in the cost function of a link, it is not obvious what functional form that cost should be in. In this case, 1/EGR$^2$ will opt towards longer, higher throughput paths, while 1/EGR may punish length a bit more while still preferring higher-throughput paths than a link cost of 1.

\subsection{Results}
\label{subsec:routingres}

 Figure \ref{fig:costcomparison} plots distillable entanglement achieved with paths found using a Dijkstra's weighted algorithm with appropriate link costs, as well as the distillable entanglement achieved using paths found by a time-intensive exhaustive path search. In the limit of perfect gate and channel fidelities, we find that Dijkstra's weighted algorithm with a link cost of 1/EGR$^2$ will slightly outperform a link cost of 1/EGR, while the inverse is true for lower fidelities across the network. A cost of 1/EGR$^2$ is more selective for higher channel EGRs to the point where longer paths become more attractive if they allow for increased overall flow (such as in the high-gate high-fidelity case). A cost of 1/EGR is less selective towards higher channel EGRs and, comparatively, more selective toward shorter path lengths. However, we find that, in the limit of low gate and channel fidelities, there is little difference between the performance of our exhaustive path search and a weighted algorithm with a link cost of 1/EGR or 1/EGR$^2$. For all network parameters, a link cost of $1$ did not outperform any of the costs that consider entanglement generation. Our work shows that the distillable entanglement achieved in a path found using Dijkstra's weighted algorithm comes very close to that found using our exhaustive path search, particularly if an effective link cost is utilized.  We also see that the inclusion of imperfect gate fidelities in our calculations results in significantly reducing our rates across the network from those achieved in Figure \ref{fig:costcomparison}a) where the gate fidelity equals 1, to Figure \ref{fig:costcomparison}b) where the gate fidelity equals 0.99, motivating the need to account for processor imperfections in quantum network models. Figure \ref{fig:trianglepaths} depicts a sample of paths chosen using the three different path search methods for a network of perfect gate and measurement fidelities, and channel fidelities of 0.99. Note that our weighted algorithms did not include component fidelities as link cost inputs.

\section{Multi-Path Routing in Differing Quantum Network Topologies}
\label{sec:topk}
A quantum network with a rich multipath topology allows an increase of entanglement generation over a standard quantum repeater chain by exploiting the multiple available paths between source and destination. Choosing the optimal topology for a network is an open question, as optimal use of the topologies depends on application and resource constraints \cite{dowling2018networktopology, chaves2020networktopology}. In this section, we aim to build a framework through which to properly evaluate network topology in the context of purification on bipartite states.

By comparing a square-, hexagonal-,  and triangular- grid topology, we can explore the dependence of distillable entanglement on node connectedness and resource constraints. We use a protocol for multi-path routing, where overall entanglement between two nodes can be increased by distributing entangled pairs over as many disjoint paths as are available in the network. While real quantum networks will not necessarily follow such strict grid-like forms in their layout, we can use these basic topologies as models to explore the trade-offs of sparsely vs. densely connected graphs by demonstrating how entanglement generation rates scale in each architecture.

\label{sec:multiroute}
\begin{figure}
  \centering
  \includegraphics[width=0.9\linewidth]{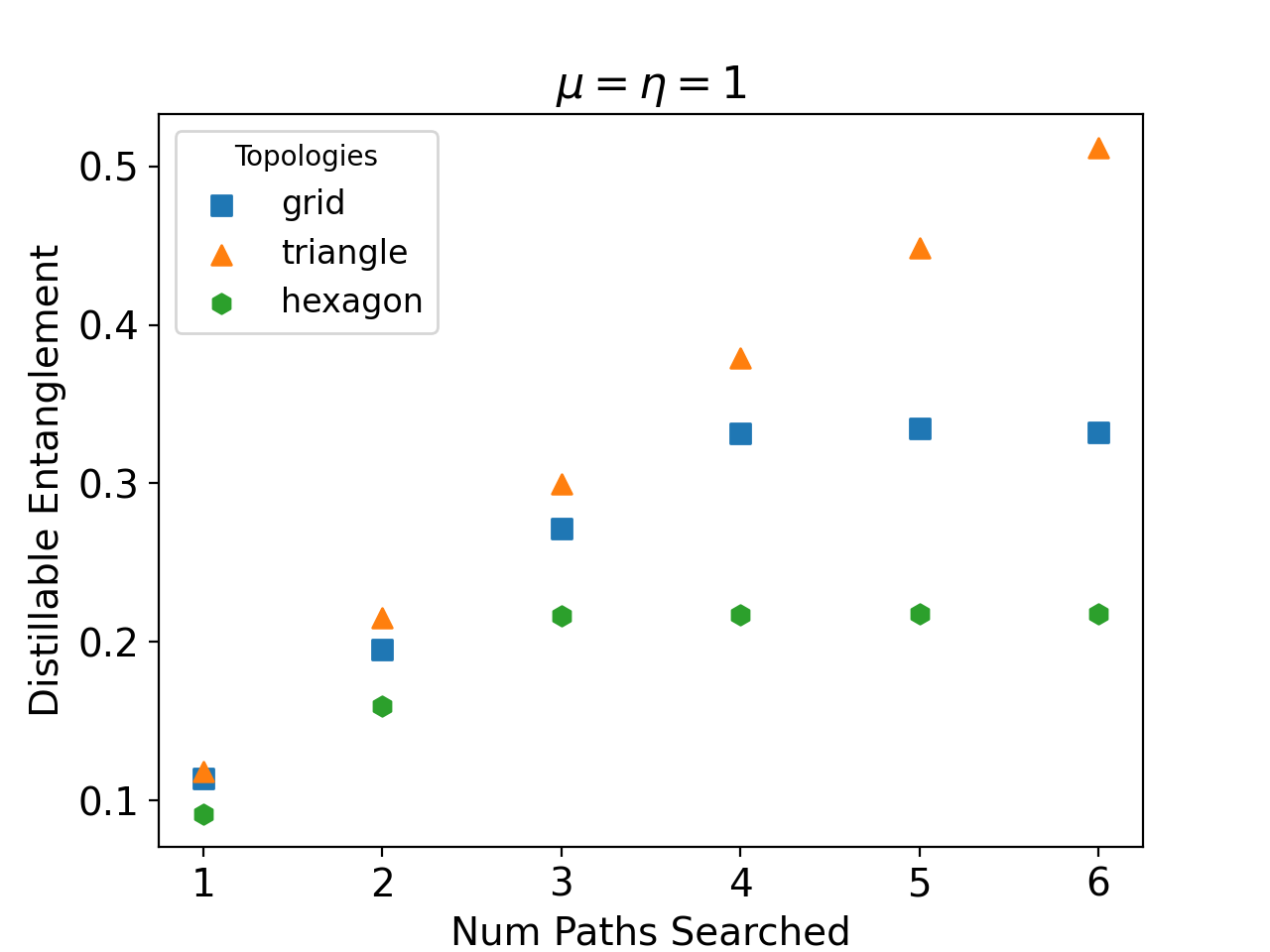}
  \centering
  \includegraphics[width=0.9\linewidth]{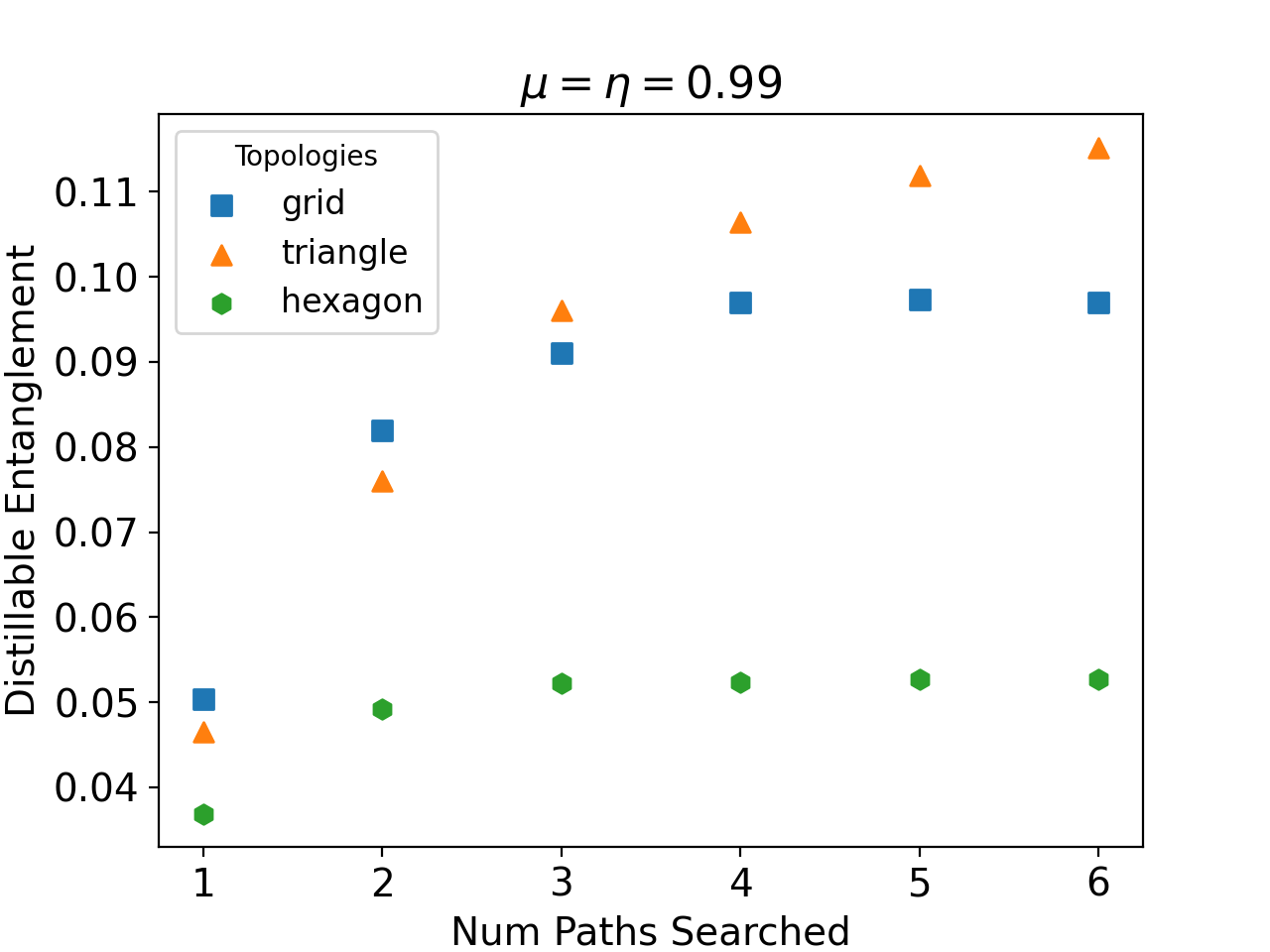}
  \caption{Distillable entanglement (normalized by average channel EGR) with equivalent channel EGR spread of 8-32 across topologies; Gate and measurement fidelity = 1 (top) and 0.99 (bottom), channel fidelity = 0.91, hop-by-hop distance = 4. The $x$-axis represents the number of paths simultaneously generating entanglement across the network. Note the difference in scale between the y-axes.}
  \label{fig:equalchannelEGR}
\end{figure}

\begin{figure}
    \includegraphics[width=0.9\linewidth]{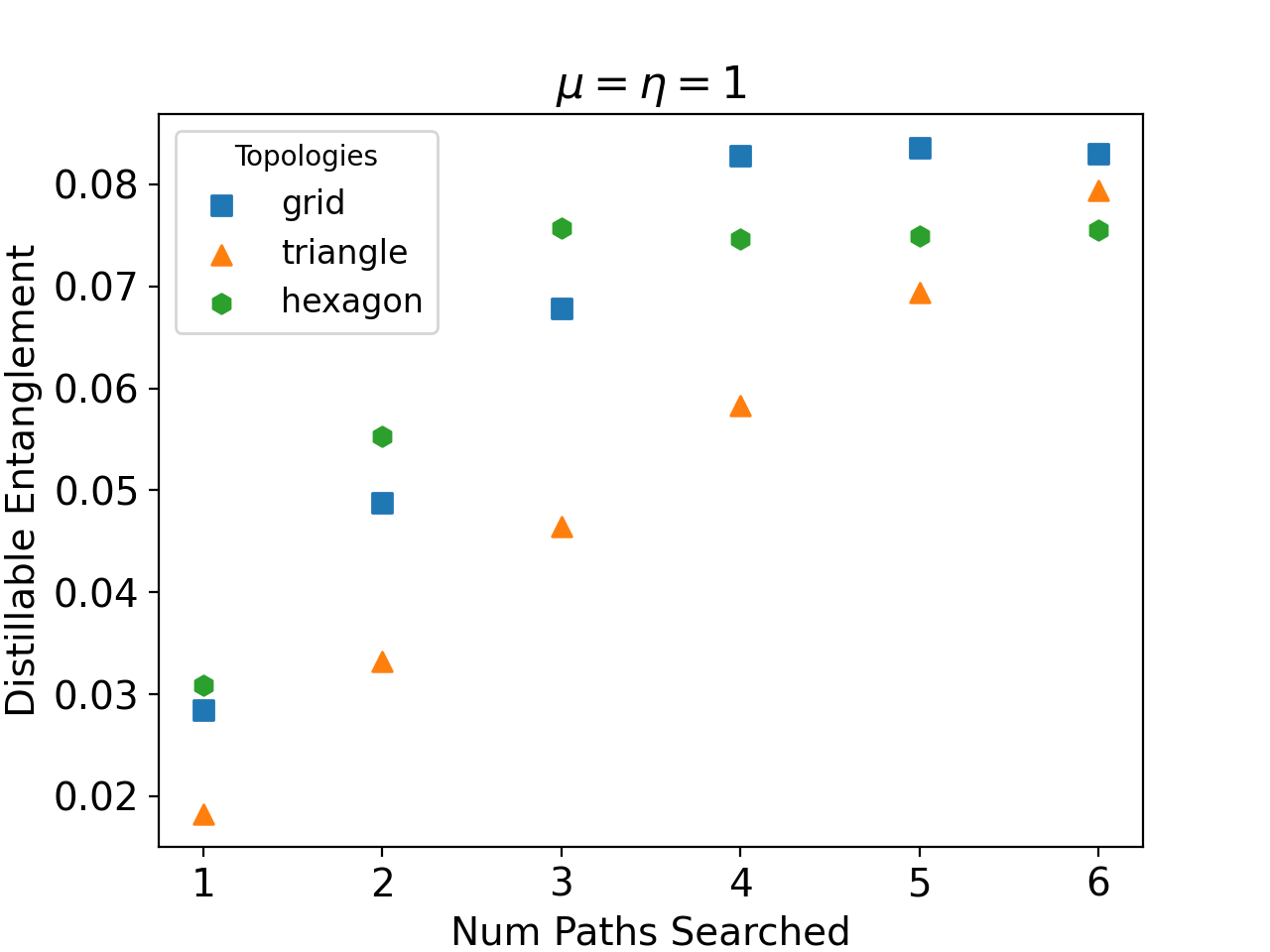}
    \includegraphics[width=0.9\linewidth]{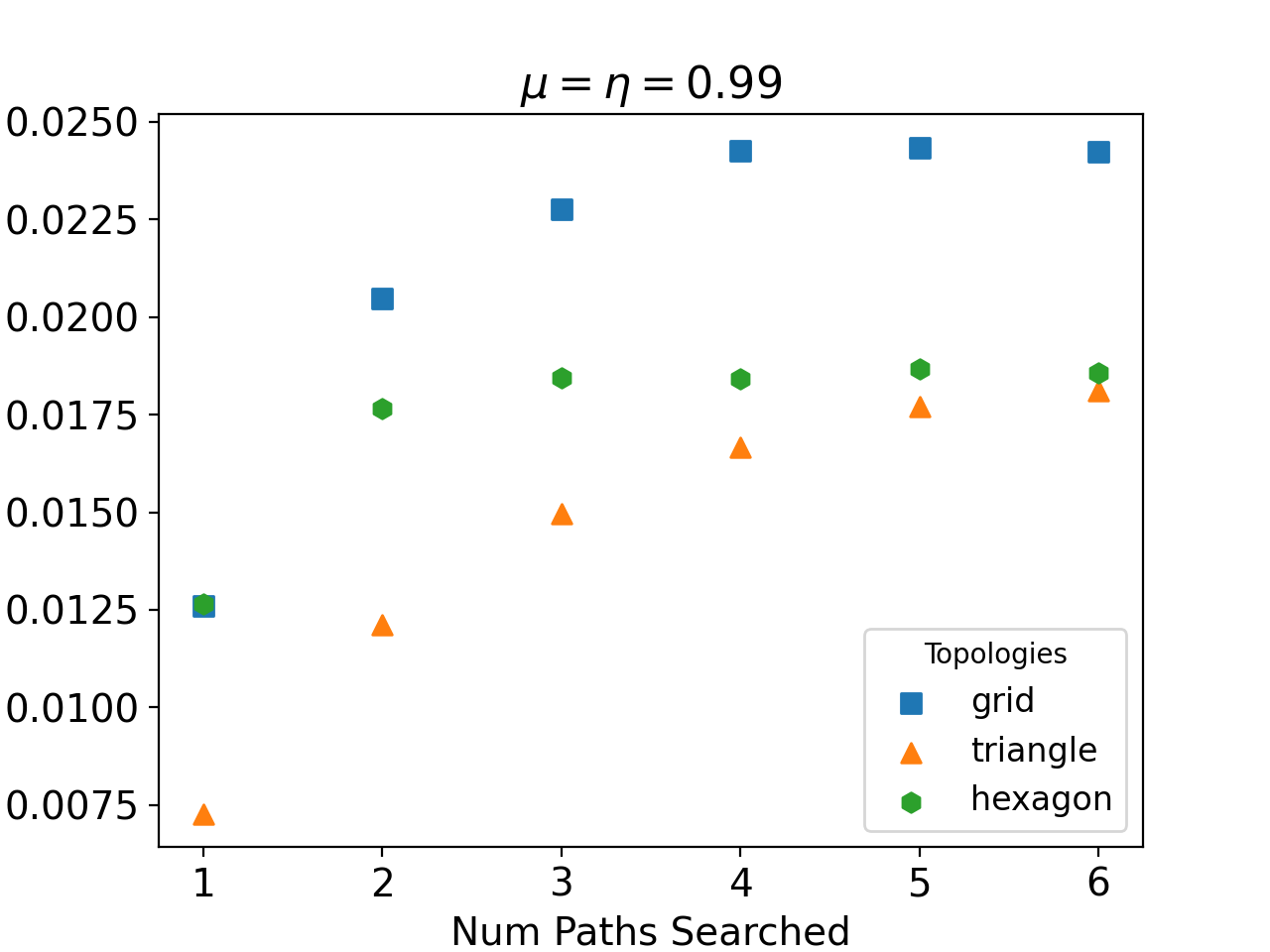}
  \caption{Distillable entanglement (normalized by average repeater EGR) with equivalent repeater EGR of 16-128 across topologies; gate and measurement fidelity = 1 (top) and 0.99 (bottom), channel fidelity = 0.91, and hop-by-hop distance = 4. The $x$-axis represents the number of paths simultaneously generating entanglement across the network. Note the difference in scale between the y-axes.}
  \label{fig:equalnodeEGR}
\end{figure}

\subsection{Multipath Routing of a Single Entanglement Flow}
\label{subsec:multipath}

Recently, Guha et al. \cite{saikatrouting} introduced the idea of simultaneously routing through multiple paths in order to achieve large gains in entanglement rates over the rate which can be achieved over a single chain of quantum repeaters. They defined a greedy algorithm that searches for the $k$ shortest edge-disjoint paths available in the network where $k$ is limited by the number of edge-disjoint paths available. They then attempt entanglement generation across each of these paths simultaneously. The total distillable entanglement that is achieved is the sum across the paths. Here, we use a similar greedy algorithm, replacing our search for shortest edge-disjoint paths with a search for shortest weighted edge-disjoint paths, using our link cost of 1/EGR from Section \ref{sec:routing}. This was selected due to comparable performance with an exhaustive path search (particularly in the case of imperfect gate fidelities), as shown in Figure \ref{fig:costcomparison}. A multi-path routing scheme strongly motivates the need for reducing an exhaustive path search, as the computational time grows with the number of paths available in the network.

In order to adequately compare networks and the achievable distillable entanglement, we introduce the definition of \emph{repeater EGR}. While channel EGR represents the number of entangled pairs generated across a single channel, repeater EGR represents the summation of all purified pairs a repeater shares with its neighbors. Maintaining repeater EGR across topologies translates a 2:3:4 ratio for channel EGR in a triangular, square, and hexagonal lattice, respectively, i.e., an average channel in the triangular network will have half the EGR of an average channel in the hexagonal network, in order to maintain the same repeater EGR. We can use both channel EGR and repeater EGR as a tool for comparing resource constraints on the topologies. 

\vspace{5mm}

\subsection{Results}
\label{subsec:multirouteres}
 To determine the trade-offs of topology, paths, and bandwidth, we compare distillable entanglement using the two equivalent resource frameworks of channel EGR and repeater EGR. We demonstrate the advantage of multi-path routing by plotting the distillable entanglement against the number of paths through which we simultaneously route entanglement.

Figure \ref{fig:equalchannelEGR} plots distillable entanglement in networks where all networks have an equivalently uniform spread of channel EGR. For all numbers of paths through which entanglement is simultaneously routed, we see that a triangular network maximizes distillable entanglement shared between two repeaters. This is a result of the increased node degree, as an increased number of paths available between repeaters leads to a) more path options for routing, and b) greater max EGR that can be shared between repeaters. We can relate this increased path availability to an increased percolation threshold \cite{sykes1964percolation}, which defines the minimum probability of link formation that allows for a giant connected component to exist within a network. Out of the three lattice types we explore, a triangle has the minimum percolation threshold, meaning that it can withstand connectivity at lower link probabilities than the other two lattices.

Figure \ref{fig:equalnodeEGR} plots distillable entanglement for three different networks with equivalent repeater EGR, as defined in Section \ref{subsec:multipath}.  A hexagonal lattice will be able to maximize bandwidth across a single path, as EGR across each channel is 4/3 the rate of a channel in a square lattice. However, in a multi-path situation, a square grid outperforms both the triangular and the hexagonal lattice. This can be explained intuitively upon a closer examination of the network topologies. While a triangular lattice offers the greatest number of disjoint paths between repeaters, the paths found in the later iterations of our greedy algorithm will likely be much longer than the shortest paths found in the beginning stages. As channel imperfections across longer paths lead to exponential decrease of fidelity, our overall distillable entanglement gained from these extra paths will decrease to a marginal amount, as shown. A hexagonal lattice suffers from a similar path length imbalance between repeaters, with even fewer overall path options.

Here, we have compared performance of multi-path entanglement routing between quantum topologies using two different resource equivalence frameworks: 1) equivalent channel EGR and 2) equivalent repeater EGR. For both frameworks we compared distillable entanglement achieved over the same hop-by-hop separation for two different sets of network parameters. By using this framework, we motivate the need to consider resource constraints and network size when designing topologies for quantum networks. We demonstrate that, while greater node degree allows for increased number of paths between nodes, increased path availability is only beneficial if the length of those paths is adequately constrained.

Additionally, we may want to consider alternative routing algorithms for improved performance. For example, multi-path routing where paths are not edge-disjoint may lead to improved resource sharing along edges with higher entanglement generation rates. Physically, this could represent separate entanglement flows that are handled individually but on overlapping channels. While this would be an added layer of repeater functionality currently not handled in our model, developing processors capable of handling multiple separate entanglement flows across one channel may result in increased overall distillable entanglement on a network.

\section{Discussion and Conclusions} 
\label{sec:disc}

In this \emph{Article} we have analyzed routing and purification protocols for distributing entanglement among source-destination pairs. We have extensively discussed our treatment of purification routines to be performed on repeater chains of varied entanglement generation across links with limited memory decoherence times. This was followed by a presentation of different network parameter regimes to determine different routing schemes, and a discussion of appropriate link costs for Dijkstra's weighted path algorithm. Finally, we finished with a analysis of how a network can provide increased entanglement rates by using multi-path routing between repeaters. The approach and calculations presented here provide a baseline for routing on near-term quantum networks with resource and performance constraints on individual repeaters, as well as the consideration of varied purification schemes optimized for channel bandwidth and repeater gate operation fidelity. We have used distillable entanglement as an appropriate optimization metric for maximizing information flow across a quantum network.


This work opens up a number of new questions for future exploration. Even in a simplified model where entanglement generation is deterministic and coherence time is finite, we demonstrate the need to tune distillation protocols across different channels of the network in order to distribute information at the highest rate. Accounting for dynamic noise models will allow us to better optimize purification and memory storage time for a synchronized network continuously generating entanglement. In addition, by decomposing our entanglement generation rate into probabilistic source rates and sizing of qubit buffers on individual repeaters, we will be able to explore optimal architectures for repeaters given the network topology and overall figures of merit.

Quantum networks are critical components for the development of large scale quantum systems. Emerging quantum applications will drive new requirements on the underlying quantum network devices and protocols and fundamental to these future quantum networks is entanglement routing. Looking ahead, the development of constraint-based entanglement routing protocols based on topology, noise models, link costs, resource usage, and diverse quantum repeater architectures will be crucial in overcoming the limitations of near-term quantum networks. In this \emph{Article}, we present the groundwork for a working near-term quantum internet built from repeaters with limited processing fidelity. These results are the foundation for the community to build future quantum networking developments on. 

\begin{acknowledgments}
The authors thank Prof. Don Towsley (University of Massachusetts), Prof. Saikat Guha (University of Arizona), and Dr. Michael Fanto (AFRL) for insightful discussions and comments on the work. Michelle Victora is partially supported by the U.S. Air Force under FA8750-20-P-1721. Stefan Krastanov is partially supported under the U.S. Department of Energy, Office of Science, Basic Energy  Sciences  (BES),  Materials  Sciences  and  Engineering Division under FWP ERKCK47. Simulations made use of the NetworkX program for analysing networks in Python \cite{swart2008networkx}.

\end{acknowledgments}

\bibliography{Routing.bib}

\begin{thebibliography}{49}%
\makeatletter
\providecommand \@ifxundefined [1]{%
 \@ifx{#1\undefined}
}%
\providecommand \@ifnum [1]{%
 \ifnum #1\expandafter \@firstoftwo
 \else \expandafter \@secondoftwo
 \fi
}%
\providecommand \@ifx [1]{%
 \ifx #1\expandafter \@firstoftwo
 \else \expandafter \@secondoftwo
 \fi
}%
\providecommand \natexlab [1]{#1}%
\providecommand \enquote  [1]{``#1''}%
\providecommand \bibnamefont  [1]{#1}%
\providecommand \bibfnamefont [1]{#1}%
\providecommand \citenamefont [1]{#1}%
\providecommand \href@noop [0]{\@secondoftwo}%
\providecommand \href [0]{\begingroup \@sanitize@url \@href}%
\providecommand \@href[1]{\@@startlink{#1}\@@href}%
\providecommand \@@href[1]{\endgroup#1\@@endlink}%
\providecommand \@sanitize@url [0]{\catcode `\\12\catcode `\$12\catcode
  `\&12\catcode `\#12\catcode `\^12\catcode `\_12\catcode `\%12\relax}%
\providecommand \@@startlink[1]{}%
\providecommand \@@endlink[0]{}%
\providecommand \url  [0]{\begingroup\@sanitize@url \@url }%
\providecommand \@url [1]{\endgroup\@href {#1}{\urlprefix }}%
\providecommand \urlprefix  [0]{URL }%
\providecommand \Eprint [0]{\href }%
\providecommand \doibase [0]{http://dx.doi.org/}%
\providecommand \selectlanguage [0]{\@gobble}%
\providecommand \bibinfo  [0]{\@secondoftwo}%
\providecommand \bibfield  [0]{\@secondoftwo}%
\providecommand \translation [1]{[#1]}%
\providecommand \BibitemOpen [0]{}%
\providecommand \bibitemStop [0]{}%
\providecommand \bibitemNoStop [0]{.\EOS\space}%
\providecommand \EOS [0]{\spacefactor3000\relax}%
\providecommand \BibitemShut  [1]{\csname bibitem#1\endcsname}%
\let\auto@bib@innerbib\@empty
\bibitem [{\citenamefont {Van~Meter}(2014)}]{vanmeternetworking}%
  \BibitemOpen
  \bibfield  {author} {\bibinfo {author} {\bibfnamefont {R.}~\bibnamefont
  {Van~Meter}},\ }\href@noop {} {\emph {\bibinfo {title} {Quantum
  networking}}}\ (\bibinfo  {publisher} {John Wiley \& Sons},\ \bibinfo {year}
  {2014})\BibitemShut {NoStop}%
\bibitem [{\citenamefont {Kimble}(2008)}]{kimbleinternet}%
  \BibitemOpen
  \bibfield  {author} {\bibinfo {author} {\bibfnamefont {H.~J.}\ \bibnamefont
  {Kimble}},\ }\href {https://www.nature.com/articles/nature07127} {\bibfield
  {journal} {\bibinfo  {journal} {Nature}\ }\textbf {\bibinfo {volume} {453}},\
  \bibinfo {pages} {1023} (\bibinfo {year} {2008})}\BibitemShut {NoStop}%
\bibitem [{\citenamefont {Castelvecchi}(2018)}]{castelvecchiinternet}%
  \BibitemOpen
  \bibfield  {author} {\bibinfo {author} {\bibfnamefont {D.}~\bibnamefont
  {Castelvecchi}},\ }\href {https://www.nature.com/articles/d41586-018-01835-3}
  {\bibfield  {journal} {\bibinfo  {journal} {Nature}\ }\textbf {\bibinfo
  {volume} {554}},\ \bibinfo {pages} {289} (\bibinfo {year}
  {2018})}\BibitemShut {NoStop}%
\bibitem [{\citenamefont {Wehner}\ \emph {et~al.}(2018)\citenamefont {Wehner},
  \citenamefont {Elkouss},\ and\ \citenamefont {Hanson}}]{wehnerinternet}%
  \BibitemOpen
  \bibfield  {author} {\bibinfo {author} {\bibfnamefont {S.}~\bibnamefont
  {Wehner}}, \bibinfo {author} {\bibfnamefont {D.}~\bibnamefont {Elkouss}}, \
  and\ \bibinfo {author} {\bibfnamefont {R.}~\bibnamefont {Hanson}},\ }\href
  {https://science.sciencemag.org/content/362/6412/eaam9288} {\bibfield
  {journal} {\bibinfo  {journal} {Nature}\ }\textbf {\bibinfo {volume} {362}}
  (\bibinfo {year} {2018})}\BibitemShut {NoStop}%
\bibitem [{\citenamefont {Broadbent}\ and\ \citenamefont
  {Schaffner}(2016)}]{beyondqkd}%
  \BibitemOpen
  \bibfield  {author} {\bibinfo {author} {\bibfnamefont {A.}~\bibnamefont
  {Broadbent}}\ and\ \bibinfo {author} {\bibfnamefont {C.}~\bibnamefont
  {Schaffner}},\ }\href {https://link.springer.com/article/10.1007} {\bibfield
  {journal} {\bibinfo  {journal} {Designs, Codes and Cryptography}\ }\textbf
  {\bibinfo {volume} {78}},\ \bibinfo {pages} {351} (\bibinfo {year}
  {2016})}\BibitemShut {NoStop}%
\bibitem [{dis(2013)}]{distributedqc2}%
  \BibitemOpen
  \href {\doibase 10.1098/rspa.2012.0686} {\emph {\bibinfo {title} {Efficient
  Distributed Quantum Computing}}},\ Vol.\ \bibinfo {volume} {469}\ (\bibinfo
  {year} {2013})\BibitemShut {NoStop}%
\bibitem [{\citenamefont {Ge}\ \emph {et~al.}(2018)\citenamefont {Ge},
  \citenamefont {Jacobs}, \citenamefont {Eldredge}, \citenamefont {Gorshkov},\
  and\ \citenamefont {Foss-Feig}}]{distributedqc3}%
  \BibitemOpen
  \bibfield  {author} {\bibinfo {author} {\bibfnamefont {W.}~\bibnamefont
  {Ge}}, \bibinfo {author} {\bibfnamefont {K.}~\bibnamefont {Jacobs}}, \bibinfo
  {author} {\bibfnamefont {Z.}~\bibnamefont {Eldredge}}, \bibinfo {author}
  {\bibfnamefont {A.~V.}\ \bibnamefont {Gorshkov}}, \ and\ \bibinfo {author}
  {\bibfnamefont {M.}~\bibnamefont {Foss-Feig}},\ }\href {\doibase
  10.1103/PhysRevLett.121.043604} {\bibfield  {journal} {\bibinfo  {journal}
  {Physical Review Letters}\ }\textbf {\bibinfo {volume} {121}},\ \bibinfo
  {pages} {043604} (\bibinfo {year} {2018})}\BibitemShut {NoStop}%
\bibitem [{\citenamefont {Bennett}\ and\ \citenamefont
  {Brassard}(2014)}]{multipartysecrets}%
  \BibitemOpen
  \bibfield  {author} {\bibinfo {author} {\bibfnamefont {C.~H.}\ \bibnamefont
  {Bennett}}\ and\ \bibinfo {author} {\bibfnamefont {G.}~\bibnamefont
  {Brassard}},\ }\href {https://doi.org/10.1016/j.tcs.2014.05.025} {\bibfield
  {journal} {\bibinfo  {journal} {Theoretical Computer Science}\ }\textbf
  {\bibinfo {volume} {560}},\ \bibinfo {pages} {7} (\bibinfo {year}
  {2014})}\BibitemShut {NoStop}%
\bibitem [{\citenamefont {Eldredge}\ \emph {et~al.}(2018)\citenamefont
  {Eldredge}, \citenamefont {Foss-Feig}, \citenamefont {Gross}, \citenamefont
  {Rolston},\ and\ \citenamefont {Gorshkov}}]{sensing1}%
  \BibitemOpen
  \bibfield  {author} {\bibinfo {author} {\bibfnamefont {Z.}~\bibnamefont
  {Eldredge}}, \bibinfo {author} {\bibfnamefont {M.}~\bibnamefont {Foss-Feig}},
  \bibinfo {author} {\bibfnamefont {J.~A.}\ \bibnamefont {Gross}}, \bibinfo
  {author} {\bibfnamefont {S.~L.}\ \bibnamefont {Rolston}}, \ and\ \bibinfo
  {author} {\bibfnamefont {A.~V.}\ \bibnamefont {Gorshkov}},\ }\href {\doibase
  10.1103/PhysRevA.97.042337} {\bibfield  {journal} {\bibinfo  {journal}
  {Physical Review A}\ }\textbf {\bibinfo {volume} {97}},\ \bibinfo {pages}
  {042337} (\bibinfo {year} {2018})}\BibitemShut {NoStop}%
\bibitem [{\citenamefont {Proctor}\ \emph {et~al.}(2018)\citenamefont
  {Proctor}, \citenamefont {Knott},\ and\ \citenamefont
  {Dunningham}}]{sensing2}%
  \BibitemOpen
  \bibfield  {author} {\bibinfo {author} {\bibfnamefont {T.~J.}\ \bibnamefont
  {Proctor}}, \bibinfo {author} {\bibfnamefont {P.~A.}\ \bibnamefont {Knott}},
  \ and\ \bibinfo {author} {\bibfnamefont {J.~A.}\ \bibnamefont {Dunningham}},\
  }\href {\doibase 10.1103/PhysRevLett.120.080501} {\bibfield  {journal}
  {\bibinfo  {journal} {Physical Review Letters}\ }\textbf {\bibinfo {volume}
  {120}},\ \bibinfo {pages} {080501} (\bibinfo {year} {2018})}\BibitemShut
  {NoStop}%
\bibitem [{\citenamefont {Fitzsimons}(2017)}]{blindqc}%
  \BibitemOpen
  \bibfield  {author} {\bibinfo {author} {\bibfnamefont {J.~F.}\ \bibnamefont
  {Fitzsimons}},\ }\href {https://www.nature.com/articles/s41534-017-0025-3}
  {\bibfield  {journal} {\bibinfo  {journal} {npj Quantum inf}\ }\textbf
  {\bibinfo {volume} {3}} (\bibinfo {year} {2017})}\BibitemShut {NoStop}%
\bibitem [{\citenamefont {Yin}\ \emph {et~al.}(2020)\citenamefont {Yin},
  \citenamefont {Li}, \citenamefont {Liao}, \citenamefont {Yang}, \citenamefont
  {Cao}, \citenamefont {Zhang}, \citenamefont {Ren}, \citenamefont {Cai},
  \citenamefont {Liu}, \citenamefont {Li}, \citenamefont {Shu}, \citenamefont
  {Huang}, \citenamefont {Deng}, \citenamefont {Li}, \citenamefont {Zhang},
  \citenamefont {Liu}, \citenamefont {Chen}, \citenamefont {Lu}, \citenamefont
  {Wang}, \citenamefont {Xu}, \citenamefont {Wang}, \citenamefont {Peng},
  \citenamefont {Ekert},\ and\ \citenamefont {Pan}}]{longrangeentanglement}%
  \BibitemOpen
  \bibfield  {author} {\bibinfo {author} {\bibfnamefont {J.}~\bibnamefont
  {Yin}}, \bibinfo {author} {\bibfnamefont {Y.-H.}\ \bibnamefont {Li}},
  \bibinfo {author} {\bibfnamefont {S.-K.}\ \bibnamefont {Liao}}, \bibinfo
  {author} {\bibfnamefont {M.}~\bibnamefont {Yang}}, \bibinfo {author}
  {\bibfnamefont {Y.}~\bibnamefont {Cao}}, \bibinfo {author} {\bibfnamefont
  {L.}~\bibnamefont {Zhang}}, \bibinfo {author} {\bibfnamefont {J.-G.}\
  \bibnamefont {Ren}}, \bibinfo {author} {\bibfnamefont {W.-Q.}\ \bibnamefont
  {Cai}}, \bibinfo {author} {\bibfnamefont {W.-Y.}\ \bibnamefont {Liu}},
  \bibinfo {author} {\bibfnamefont {S.-L.}\ \bibnamefont {Li}}, \bibinfo
  {author} {\bibfnamefont {R.}~\bibnamefont {Shu}}, \bibinfo {author}
  {\bibfnamefont {Y.-M.}\ \bibnamefont {Huang}}, \bibinfo {author}
  {\bibfnamefont {L.}~\bibnamefont {Deng}}, \bibinfo {author} {\bibfnamefont
  {L.}~\bibnamefont {Li}}, \bibinfo {author} {\bibfnamefont {Q.}~\bibnamefont
  {Zhang}}, \bibinfo {author} {\bibfnamefont {N.-L.}\ \bibnamefont {Liu}},
  \bibinfo {author} {\bibfnamefont {Y.-A.}\ \bibnamefont {Chen}}, \bibinfo
  {author} {\bibfnamefont {C.-Y.}\ \bibnamefont {Lu}}, \bibinfo {author}
  {\bibfnamefont {X.-B.}\ \bibnamefont {Wang}}, \bibinfo {author}
  {\bibfnamefont {F.}~\bibnamefont {Xu}}, \bibinfo {author} {\bibfnamefont
  {J.-Y.}\ \bibnamefont {Wang}}, \bibinfo {author} {\bibfnamefont {C.-Z.}\
  \bibnamefont {Peng}}, \bibinfo {author} {\bibfnamefont {A.}~\bibnamefont
  {Ekert}}, \ and\ \bibinfo {author} {\bibfnamefont {J.-W.}\ \bibnamefont
  {Pan}},\ }\href {https://www.nature.com/articles/s41586-020-2401-y}
  {\bibfield  {journal} {\bibinfo  {journal} {Nature}\ }\textbf {\bibinfo
  {volume} {582}},\ \bibinfo {pages} {501} (\bibinfo {year}
  {2020})}\BibitemShut {NoStop}%
\bibitem [{\citenamefont {Yin}\ \emph {et~al.}(2017)\citenamefont {Yin},
  \citenamefont {Cao}, \citenamefont {Li}, \citenamefont {Liao}, \citenamefont
  {Zhang}, \citenamefont {Ren}, \citenamefont {Cai}, \citenamefont {Liu},
  \citenamefont {Li}, \citenamefont {Dai}, \citenamefont {Li}, \citenamefont
  {Lu}, \citenamefont {Gong}, \citenamefont {Xu}, \citenamefont {Li},
  \citenamefont {Li}, \citenamefont {Yin}, \citenamefont {Jiang}, \citenamefont
  {Li}, \citenamefont {Jia}, \citenamefont {Ren}, \citenamefont {He},
  \citenamefont {Zhou}, \citenamefont {Zhang}, \citenamefont {Wang},
  \citenamefont {Chang}, \citenamefont {Zhu}, \citenamefont {Liu},
  \citenamefont {Chen}, \citenamefont {Lu}, \citenamefont {Shu}, \citenamefont
  {Peng}, \citenamefont {Wang},\ and\ \citenamefont
  {Pan}}]{satelliteentanglement1}%
  \BibitemOpen
  \bibfield  {author} {\bibinfo {author} {\bibfnamefont {J.}~\bibnamefont
  {Yin}}, \bibinfo {author} {\bibfnamefont {Y.}~\bibnamefont {Cao}}, \bibinfo
  {author} {\bibfnamefont {Y.-H.}\ \bibnamefont {Li}}, \bibinfo {author}
  {\bibfnamefont {S.-K.}\ \bibnamefont {Liao}}, \bibinfo {author}
  {\bibfnamefont {L.}~\bibnamefont {Zhang}}, \bibinfo {author} {\bibfnamefont
  {J.-G.}\ \bibnamefont {Ren}}, \bibinfo {author} {\bibfnamefont {W.-Q.}\
  \bibnamefont {Cai}}, \bibinfo {author} {\bibfnamefont {W.-Y.}\ \bibnamefont
  {Liu}}, \bibinfo {author} {\bibfnamefont {B.}~\bibnamefont {Li}}, \bibinfo
  {author} {\bibfnamefont {H.}~\bibnamefont {Dai}}, \bibinfo {author}
  {\bibfnamefont {G.-B.}\ \bibnamefont {Li}}, \bibinfo {author} {\bibfnamefont
  {Q.-M.}\ \bibnamefont {Lu}}, \bibinfo {author} {\bibfnamefont {Y.~H.}\
  \bibnamefont {Gong}}, \bibinfo {author} {\bibfnamefont {Y.}~\bibnamefont
  {Xu}}, \bibinfo {author} {\bibfnamefont {S.-L.}\ \bibnamefont {Li}}, \bibinfo
  {author} {\bibfnamefont {F.-Z.}\ \bibnamefont {Li}}, \bibinfo {author}
  {\bibfnamefont {Y.~Y.}\ \bibnamefont {Yin}}, \bibinfo {author} {\bibfnamefont
  {Z.-Q.}\ \bibnamefont {Jiang}}, \bibinfo {author} {\bibfnamefont
  {M.}~\bibnamefont {Li}}, \bibinfo {author} {\bibfnamefont {J.-J.}\
  \bibnamefont {Jia}}, \bibinfo {author} {\bibfnamefont {G.}~\bibnamefont
  {Ren}}, \bibinfo {author} {\bibfnamefont {D.}~\bibnamefont {He}}, \bibinfo
  {author} {\bibfnamefont {Y.-L.}\ \bibnamefont {Zhou}}, \bibinfo {author}
  {\bibfnamefont {X.-X.}\ \bibnamefont {Zhang}}, \bibinfo {author}
  {\bibfnamefont {N.}~\bibnamefont {Wang}}, \bibinfo {author} {\bibfnamefont
  {X.}~\bibnamefont {Chang}}, \bibinfo {author} {\bibfnamefont {Z.-C.}\
  \bibnamefont {Zhu}}, \bibinfo {author} {\bibfnamefont {N.-L.}\ \bibnamefont
  {Liu}}, \bibinfo {author} {\bibfnamefont {Y.-A.}\ \bibnamefont {Chen}},
  \bibinfo {author} {\bibfnamefont {C.-Y.}\ \bibnamefont {Lu}}, \bibinfo
  {author} {\bibfnamefont {R.}~\bibnamefont {Shu}}, \bibinfo {author}
  {\bibfnamefont {C.-Z.}\ \bibnamefont {Peng}}, \bibinfo {author}
  {\bibfnamefont {J.-Y.}\ \bibnamefont {Wang}}, \ and\ \bibinfo {author}
  {\bibfnamefont {J.-W.}\ \bibnamefont {Pan}},\ }\href
  {https://science.sciencemag.org/content/356/6343/1140} {\bibfield  {journal}
  {\bibinfo  {journal} {Science}\ }\textbf {\bibinfo {volume} {356}},\ \bibinfo
  {pages} {1140} (\bibinfo {year} {2017})}\BibitemShut {NoStop}%
\bibitem [{\citenamefont {J.-G.}\ \emph {et~al.}(2017)\citenamefont {J.-G.},
  \citenamefont {Xu}, \citenamefont {Yong}, \citenamefont {Zhang},
  \citenamefont {Liao}, \citenamefont {Yin}, \citenamefont {Liu}, \citenamefont
  {Cai}, \citenamefont {Yang}, \citenamefont {Li}, \citenamefont {Yang},
  \citenamefont {Han}, \citenamefont {Yao}, \citenamefont {Li}, \citenamefont
  {Wu}, \citenamefont {Wan}, \citenamefont {Liu}, \citenamefont {Liu},
  \citenamefont {Kuang}, \citenamefont {He}, \citenamefont {Shang},
  \citenamefont {Guo}, \citenamefont {Zheng}, \citenamefont {Tian},
  \citenamefont {Zhu}, \citenamefont {Liu}, \citenamefont {Lu}, \citenamefont
  {Shu}, \citenamefont {Chen}, \citenamefont {Peng}, \citenamefont {Wang},\
  and\ \citenamefont {Pan}}]{satelliteentanglement2}%
  \BibitemOpen
  \bibfield  {author} {\bibinfo {author} {\bibfnamefont {R.}~\bibnamefont
  {J.-G.}}, \bibinfo {author} {\bibfnamefont {P.}~\bibnamefont {Xu}}, \bibinfo
  {author} {\bibfnamefont {H.-L.}\ \bibnamefont {Yong}}, \bibinfo {author}
  {\bibfnamefont {L.}~\bibnamefont {Zhang}}, \bibinfo {author} {\bibfnamefont
  {S.-K.}\ \bibnamefont {Liao}}, \bibinfo {author} {\bibfnamefont
  {J.}~\bibnamefont {Yin}}, \bibinfo {author} {\bibfnamefont {W.-Y.}\
  \bibnamefont {Liu}}, \bibinfo {author} {\bibfnamefont {W.-Q.}\ \bibnamefont
  {Cai}}, \bibinfo {author} {\bibfnamefont {M.}~\bibnamefont {Yang}}, \bibinfo
  {author} {\bibfnamefont {L.}~\bibnamefont {Li}}, \bibinfo {author}
  {\bibfnamefont {K.-X.}\ \bibnamefont {Yang}}, \bibinfo {author}
  {\bibfnamefont {X.}~\bibnamefont {Han}}, \bibinfo {author} {\bibfnamefont
  {Y.-Q.}\ \bibnamefont {Yao}}, \bibinfo {author} {\bibfnamefont
  {J.}~\bibnamefont {Li}}, \bibinfo {author} {\bibfnamefont {H.-Y.}\
  \bibnamefont {Wu}}, \bibinfo {author} {\bibfnamefont {S.}~\bibnamefont
  {Wan}}, \bibinfo {author} {\bibfnamefont {L.}~\bibnamefont {Liu}}, \bibinfo
  {author} {\bibfnamefont {D.-Q.}\ \bibnamefont {Liu}}, \bibinfo {author}
  {\bibfnamefont {Y.-W.}\ \bibnamefont {Kuang}}, \bibinfo {author}
  {\bibfnamefont {Z.-P.}\ \bibnamefont {He}}, \bibinfo {author} {\bibfnamefont
  {P.}~\bibnamefont {Shang}}, \bibinfo {author} {\bibfnamefont
  {C.}~\bibnamefont {Guo}}, \bibinfo {author} {\bibfnamefont {R.-H.}\
  \bibnamefont {Zheng}}, \bibinfo {author} {\bibfnamefont {K.}~\bibnamefont
  {Tian}}, \bibinfo {author} {\bibfnamefont {Z.-C.}\ \bibnamefont {Zhu}},
  \bibinfo {author} {\bibfnamefont {N.-L.}\ \bibnamefont {Liu}}, \bibinfo
  {author} {\bibfnamefont {C.-Y.}\ \bibnamefont {Lu}}, \bibinfo {author}
  {\bibfnamefont {R.}~\bibnamefont {Shu}}, \bibinfo {author} {\bibfnamefont
  {Y.-A.}\ \bibnamefont {Chen}}, \bibinfo {author} {\bibfnamefont {C.-Z.}\
  \bibnamefont {Peng}}, \bibinfo {author} {\bibfnamefont {J.-Y.}\ \bibnamefont
  {Wang}}, \ and\ \bibinfo {author} {\bibfnamefont {J.-W.}\ \bibnamefont
  {Pan}},\ }\href {https://doi.org/10.1038/nature23675} {\bibfield  {journal}
  {\bibinfo  {journal} {Nature}\ }\textbf {\bibinfo {volume} {549}},\ \bibinfo
  {pages} {70} (\bibinfo {year} {2017})}\BibitemShut {NoStop}%
\bibitem [{\citenamefont {Dabbar}(2020)}]{quantuminternetishere}%
  \BibitemOpen
  \bibfield  {author} {\bibinfo {author} {\bibfnamefont {P.}~\bibnamefont
  {Dabbar}},\ }\href
  {https://www.energy.gov/articles/quantum-internet-future-here} {\enquote
  {\bibinfo {title} {The quantum internet of the future is here},}\ } (\bibinfo
  {year} {2020})\BibitemShut {NoStop}%
\bibitem [{\citenamefont {Sangouard}\ \emph {et~al.}(2011)\citenamefont
  {Sangouard}, \citenamefont {Simon}, \citenamefont {de~Riedmatten},\ and\
  \citenamefont {Gisin}}]{repeaters2}%
  \BibitemOpen
  \bibfield  {author} {\bibinfo {author} {\bibfnamefont {N.}~\bibnamefont
  {Sangouard}}, \bibinfo {author} {\bibfnamefont {C.}~\bibnamefont {Simon}},
  \bibinfo {author} {\bibfnamefont {H.}~\bibnamefont {de~Riedmatten}}, \ and\
  \bibinfo {author} {\bibfnamefont {N.}~\bibnamefont {Gisin}},\ }\href
  {\doibase 10.1103/RevModPhys.83.33} {\bibfield  {journal} {\bibinfo
  {journal} {Review of Modern Physics}\ }\textbf {\bibinfo {volume} {83}},\
  \bibinfo {pages} {33} (\bibinfo {year} {2011})}\BibitemShut {NoStop}%
\bibitem [{\citenamefont {Briegel}\ \emph {et~al.}(1998)\citenamefont
  {Briegel}, \citenamefont {D\"ur}, \citenamefont {Cirac},\ and\ \citenamefont
  {Zoller}}]{repeaters1}%
  \BibitemOpen
  \bibfield  {author} {\bibinfo {author} {\bibfnamefont {H.-J.}\ \bibnamefont
  {Briegel}}, \bibinfo {author} {\bibfnamefont {W.}~\bibnamefont {D\"ur}},
  \bibinfo {author} {\bibfnamefont {J.~I.}\ \bibnamefont {Cirac}}, \ and\
  \bibinfo {author} {\bibfnamefont {P.}~\bibnamefont {Zoller}},\ }\href
  {\doibase 10.1103/PhysRevLett.81.5932} {\bibfield  {journal} {\bibinfo
  {journal} {Physical Review Letters}\ }\textbf {\bibinfo {volume} {81}},\
  \bibinfo {pages} {5932} (\bibinfo {year} {1998})}\BibitemShut {NoStop}%
\bibitem [{\citenamefont {D\"ur}\ \emph {et~al.}(1999)\citenamefont {D\"ur},
  \citenamefont {Briegel}, \citenamefont {Cirac},\ and\ \citenamefont
  {Zoller}}]{PhysRevA.59.169}%
  \BibitemOpen
  \bibfield  {author} {\bibinfo {author} {\bibfnamefont {W.}~\bibnamefont
  {D\"ur}}, \bibinfo {author} {\bibfnamefont {H.-J.}\ \bibnamefont {Briegel}},
  \bibinfo {author} {\bibfnamefont {J.~I.}\ \bibnamefont {Cirac}}, \ and\
  \bibinfo {author} {\bibfnamefont {P.}~\bibnamefont {Zoller}},\ }\href
  {\doibase 10.1103/PhysRevA.59.169} {\bibfield  {journal} {\bibinfo  {journal}
  {Physical Review A}\ }\textbf {\bibinfo {volume} {59}},\ \bibinfo {pages}
  {169} (\bibinfo {year} {1999})}\BibitemShut {NoStop}%
\bibitem [{\citenamefont {Pirker}\ and\ \citenamefont
  {D{\"u}r}(2019)}]{quantumnetworkstack}%
  \BibitemOpen
  \bibfield  {author} {\bibinfo {author} {\bibfnamefont {A.}~\bibnamefont
  {Pirker}}\ and\ \bibinfo {author} {\bibnamefont {D{\"u}r}},\ }\href
  {https://iopscience.iop.org/article/10.1088/1367-2630/ab05f7/meta} {\bibfield
   {journal} {\bibinfo  {journal} {New Journal of Physics}\ }\textbf {\bibinfo
  {volume} {21}} (\bibinfo {year} {2019})}\BibitemShut {NoStop}%
\bibitem [{\citenamefont {Bennett}\ \emph
  {et~al.}(1996{\natexlab{a}})\citenamefont {Bennett}, \citenamefont
  {Bernstein}, \citenamefont {Popescu},\ and\ \citenamefont
  {Schumacher}}]{bennett1996concentrating}%
  \BibitemOpen
  \bibfield  {author} {\bibinfo {author} {\bibfnamefont {C.~H.}\ \bibnamefont
  {Bennett}}, \bibinfo {author} {\bibfnamefont {H.~J.}\ \bibnamefont
  {Bernstein}}, \bibinfo {author} {\bibfnamefont {S.}~\bibnamefont {Popescu}},
  \ and\ \bibinfo {author} {\bibfnamefont {B.}~\bibnamefont {Schumacher}},\
  }\href {https://journals.aps.org/pra/abstract/10.1103/PhysRevA.53.2046}
  {\bibfield  {journal} {\bibinfo  {journal} {Physical Review A}\ }\textbf
  {\bibinfo {volume} {53}},\ \bibinfo {pages} {2046} (\bibinfo {year}
  {1996}{\natexlab{a}})}\BibitemShut {NoStop}%
\bibitem [{\citenamefont {Bennett}\ \emph {et~al.}(1993)\citenamefont
  {Bennett}, \citenamefont {Brassard}, \citenamefont {Cr\'epeau}, \citenamefont
  {Jozsa}, \citenamefont {Peres},\ and\ \citenamefont {Wootters}}]{BSM1}%
  \BibitemOpen
  \bibfield  {author} {\bibinfo {author} {\bibfnamefont {C.~H.}\ \bibnamefont
  {Bennett}}, \bibinfo {author} {\bibfnamefont {G.}~\bibnamefont {Brassard}},
  \bibinfo {author} {\bibfnamefont {C.}~\bibnamefont {Cr\'epeau}}, \bibinfo
  {author} {\bibfnamefont {R.}~\bibnamefont {Jozsa}}, \bibinfo {author}
  {\bibfnamefont {A.}~\bibnamefont {Peres}}, \ and\ \bibinfo {author}
  {\bibfnamefont {W.~K.}\ \bibnamefont {Wootters}},\ }\href {\doibase
  10.1103/PhysRevLett.70.1895} {\bibfield  {journal} {\bibinfo  {journal}
  {Physical Review Letters}\ }\textbf {\bibinfo {volume} {70}},\ \bibinfo
  {pages} {1895} (\bibinfo {year} {1993})}\BibitemShut {NoStop}%
\bibitem [{\citenamefont {\ifmmode~\dot{Z}\else \.{Z}\fi{}ukowski}\ \emph
  {et~al.}(1993)\citenamefont {\ifmmode~\dot{Z}\else \.{Z}\fi{}ukowski},
  \citenamefont {Zeilinger}, \citenamefont {Horne},\ and\ \citenamefont
  {Ekert}}]{BSM2}%
  \BibitemOpen
  \bibfield  {author} {\bibinfo {author} {\bibfnamefont {M.}~\bibnamefont
  {\ifmmode~\dot{Z}\else \.{Z}\fi{}ukowski}}, \bibinfo {author} {\bibfnamefont
  {A.}~\bibnamefont {Zeilinger}}, \bibinfo {author} {\bibfnamefont {M.~A.}\
  \bibnamefont {Horne}}, \ and\ \bibinfo {author} {\bibfnamefont {A.~K.}\
  \bibnamefont {Ekert}},\ }\href {\doibase 10.1103/PhysRevLett.71.4287}
  {\bibfield  {journal} {\bibinfo  {journal} {Physical Review Letters}\
  }\textbf {\bibinfo {volume} {71}},\ \bibinfo {pages} {4287} (\bibinfo {year}
  {1993})}\BibitemShut {NoStop}%
\bibitem [{\citenamefont {Goebel}\ \emph {et~al.}(2008)\citenamefont {Goebel},
  \citenamefont {Wagenknecht}, \citenamefont {Zhang}, \citenamefont {Chen},
  \citenamefont {Chen}, \citenamefont {Schmiedmayer},\ and\ \citenamefont
  {Pan}}]{BSM3}%
  \BibitemOpen
  \bibfield  {author} {\bibinfo {author} {\bibfnamefont {A.~M.}\ \bibnamefont
  {Goebel}}, \bibinfo {author} {\bibfnamefont {C.}~\bibnamefont {Wagenknecht}},
  \bibinfo {author} {\bibfnamefont {Q.}~\bibnamefont {Zhang}}, \bibinfo
  {author} {\bibfnamefont {Y.-A.}\ \bibnamefont {Chen}}, \bibinfo {author}
  {\bibfnamefont {K.}~\bibnamefont {Chen}}, \bibinfo {author} {\bibfnamefont
  {J.}~\bibnamefont {Schmiedmayer}}, \ and\ \bibinfo {author} {\bibfnamefont
  {J.-W.}\ \bibnamefont {Pan}},\ }\href {\doibase
  10.1103/PhysRevLett.101.080403} {\bibfield  {journal} {\bibinfo  {journal}
  {Physical Review Letters}\ }\textbf {\bibinfo {volume} {101}},\ \bibinfo
  {pages} {080403} (\bibinfo {year} {2008})}\BibitemShut {NoStop}%
\bibitem [{\citenamefont {Pant}\ \emph {et~al.}(2019)\citenamefont {Pant},
  \citenamefont {Krovi}, \citenamefont {Towsley}, \citenamefont {Tassiulas},
  \citenamefont {Jiang}, \citenamefont {Basu}, \citenamefont {Englund},\ and\
  \citenamefont {Guha}}]{saikatrouting}%
  \BibitemOpen
  \bibfield  {author} {\bibinfo {author} {\bibfnamefont {M.}~\bibnamefont
  {Pant}}, \bibinfo {author} {\bibfnamefont {H.}~\bibnamefont {Krovi}},
  \bibinfo {author} {\bibfnamefont {D.}~\bibnamefont {Towsley}}, \bibinfo
  {author} {\bibfnamefont {L.}~\bibnamefont {Tassiulas}}, \bibinfo {author}
  {\bibfnamefont {L.}~\bibnamefont {Jiang}}, \bibinfo {author} {\bibfnamefont
  {P.}~\bibnamefont {Basu}}, \bibinfo {author} {\bibfnamefont {D.}~\bibnamefont
  {Englund}}, \ and\ \bibinfo {author} {\bibfnamefont {S.}~\bibnamefont
  {Guha}},\ }\href {https://www.nature.com/articles/s41534-019-0139-x}
  {\bibfield  {journal} {\bibinfo  {journal} {npj Quantum Inf}\ }\textbf
  {\bibinfo {volume} {5}} (\bibinfo {year} {2019})}\BibitemShut {NoStop}%
\bibitem [{\citenamefont {Deutsch}\ \emph {et~al.}(1996)\citenamefont
  {Deutsch}, \citenamefont {Ekert}, \citenamefont {Jozsa}, \citenamefont
  {Macchiavello}, \citenamefont {Popescu},\ and\ \citenamefont
  {Sanpera}}]{PhysRevLett.77.2818}%
  \BibitemOpen
  \bibfield  {author} {\bibinfo {author} {\bibfnamefont {D.}~\bibnamefont
  {Deutsch}}, \bibinfo {author} {\bibfnamefont {A.}~\bibnamefont {Ekert}},
  \bibinfo {author} {\bibfnamefont {R.}~\bibnamefont {Jozsa}}, \bibinfo
  {author} {\bibfnamefont {C.}~\bibnamefont {Macchiavello}}, \bibinfo {author}
  {\bibfnamefont {S.}~\bibnamefont {Popescu}}, \ and\ \bibinfo {author}
  {\bibfnamefont {A.}~\bibnamefont {Sanpera}},\ }\href {\doibase
  10.1103/PhysRevLett.77.2818} {\bibfield  {journal} {\bibinfo  {journal}
  {Physical Review Letters}\ }\textbf {\bibinfo {volume} {77}},\ \bibinfo
  {pages} {2818} (\bibinfo {year} {1996})}\BibitemShut {NoStop}%
\bibitem [{\citenamefont {Fujii}\ and\ \citenamefont
  {Yamamoto}(2009)}]{doubleselection}%
  \BibitemOpen
  \bibfield  {author} {\bibinfo {author} {\bibfnamefont {K.}~\bibnamefont
  {Fujii}}\ and\ \bibinfo {author} {\bibfnamefont {K.}~\bibnamefont
  {Yamamoto}},\ }\href {\doibase 10.1103/PhysRevA.80.042308} {\bibfield
  {journal} {\bibinfo  {journal} {Physical Review A}\ }\textbf {\bibinfo
  {volume} {80}},\ \bibinfo {pages} {042308} (\bibinfo {year}
  {2009})}\BibitemShut {NoStop}%
\bibitem [{\citenamefont {Bratzik}\ \emph {et~al.}(2013)\citenamefont
  {Bratzik}, \citenamefont {Abruzzo}, \citenamefont {Kampermann},\ and\
  \citenamefont {Bru\ss{}}}]{PhysRevA.87.062335}%
  \BibitemOpen
  \bibfield  {author} {\bibinfo {author} {\bibfnamefont {S.}~\bibnamefont
  {Bratzik}}, \bibinfo {author} {\bibfnamefont {S.}~\bibnamefont {Abruzzo}},
  \bibinfo {author} {\bibfnamefont {H.}~\bibnamefont {Kampermann}}, \ and\
  \bibinfo {author} {\bibfnamefont {D.}~\bibnamefont {Bru\ss{}}},\ }\href
  {\doibase 10.1103/PhysRevA.87.062335} {\bibfield  {journal} {\bibinfo
  {journal} {Physical Review A}\ }\textbf {\bibinfo {volume} {87}},\ \bibinfo
  {pages} {062335} (\bibinfo {year} {2013})}\BibitemShut {NoStop}%
\bibitem [{\citenamefont {Bennett}\ \emph
  {et~al.}(1996{\natexlab{b}})\citenamefont {Bennett}, \citenamefont
  {Brassard}, \citenamefont {Popescu}, \citenamefont {Schumacher},
  \citenamefont {Smolin},\ and\ \citenamefont {Wootters}}]{PhysRevLett.76.722}%
  \BibitemOpen
  \bibfield  {author} {\bibinfo {author} {\bibfnamefont {C.~H.}\ \bibnamefont
  {Bennett}}, \bibinfo {author} {\bibfnamefont {G.}~\bibnamefont {Brassard}},
  \bibinfo {author} {\bibfnamefont {S.}~\bibnamefont {Popescu}}, \bibinfo
  {author} {\bibfnamefont {B.}~\bibnamefont {Schumacher}}, \bibinfo {author}
  {\bibfnamefont {J.~A.}\ \bibnamefont {Smolin}}, \ and\ \bibinfo {author}
  {\bibfnamefont {W.~K.}\ \bibnamefont {Wootters}},\ }\href {\doibase
  10.1103/PhysRevLett.76.722} {\bibfield  {journal} {\bibinfo  {journal}
  {Physical Review Letters}\ }\textbf {\bibinfo {volume} {76}},\ \bibinfo
  {pages} {722} (\bibinfo {year} {1996}{\natexlab{b}})}\BibitemShut {NoStop}%
\bibitem [{\citenamefont {Aschauer}(2005)}]{aschauer2005}%
  \BibitemOpen
  \bibfield  {author} {\bibinfo {author} {\bibfnamefont {H.}~\bibnamefont
  {Aschauer}},\ }\emph {\bibinfo {title} {Quantum communication in noisy
  environments}},\ \href
  {https://edoc.ub.uni-muenchen.de/3588/1/Aschauer_Hans.pdf} {Ph.D. thesis},\
  \bibinfo  {school} {lmu} (\bibinfo {year} {2005})\BibitemShut {NoStop}%
\bibitem [{\citenamefont {Krastanov}\ \emph {et~al.}(2019)\citenamefont
  {Krastanov}, \citenamefont {Albert},\ and\ \citenamefont
  {Jiang}}]{stefanpurification}%
  \BibitemOpen
  \bibfield  {author} {\bibinfo {author} {\bibfnamefont {S.}~\bibnamefont
  {Krastanov}}, \bibinfo {author} {\bibfnamefont {V.~V.}\ \bibnamefont
  {Albert}}, \ and\ \bibinfo {author} {\bibfnamefont {L.}~\bibnamefont
  {Jiang}},\ }\href {https://quantum-journal.org/papers/q-2019-02-18-123/}
  {\bibfield  {journal} {\bibinfo  {journal} {Quantum}\ }\textbf {\bibinfo
  {volume} {3}} (\bibinfo {year} {2019})}\BibitemShut {NoStop}%
\bibitem [{\citenamefont {Dehaene}\ \emph {et~al.}(2003)\citenamefont
  {Dehaene}, \citenamefont {Van~den Nest}, \citenamefont {De~Moor},\ and\
  \citenamefont {Verstraete}}]{PhysRevA.67.022310}%
  \BibitemOpen
  \bibfield  {author} {\bibinfo {author} {\bibfnamefont {J.}~\bibnamefont
  {Dehaene}}, \bibinfo {author} {\bibfnamefont {M.}~\bibnamefont {Van~den
  Nest}}, \bibinfo {author} {\bibfnamefont {B.}~\bibnamefont {De~Moor}}, \ and\
  \bibinfo {author} {\bibfnamefont {F.}~\bibnamefont {Verstraete}},\ }\href
  {\doibase 10.1103/PhysRevA.67.022310} {\bibfield  {journal} {\bibinfo
  {journal} {Physical Review A}\ }\textbf {\bibinfo {volume} {67}},\ \bibinfo
  {pages} {022310} (\bibinfo {year} {2003})}\BibitemShut {NoStop}%
\bibitem [{\citenamefont {Bombin}\ and\ \citenamefont
  {Martin-Delgado}(2005)}]{PhysRevA.72.032313}%
  \BibitemOpen
  \bibfield  {author} {\bibinfo {author} {\bibfnamefont {H.}~\bibnamefont
  {Bombin}}\ and\ \bibinfo {author} {\bibfnamefont {M.~A.}\ \bibnamefont
  {Martin-Delgado}},\ }\href {\doibase 10.1103/PhysRevA.72.032313} {\bibfield
  {journal} {\bibinfo  {journal} {Physical Review A}\ }\textbf {\bibinfo
  {volume} {72}},\ \bibinfo {pages} {032313} (\bibinfo {year}
  {2005})}\BibitemShut {NoStop}%
\bibitem [{\citenamefont {Schoute}\ \emph {et~al.}(2016)\citenamefont
  {Schoute}, \citenamefont {Mancinska}, \citenamefont {Islam}, \citenamefont
  {Kerenidis},\ and\ \citenamefont {Wehner}}]{schouteroutingtopologies}%
  \BibitemOpen
  \bibfield  {author} {\bibinfo {author} {\bibfnamefont {E.}~\bibnamefont
  {Schoute}}, \bibinfo {author} {\bibfnamefont {L.}~\bibnamefont {Mancinska}},
  \bibinfo {author} {\bibfnamefont {T.}~\bibnamefont {Islam}}, \bibinfo
  {author} {\bibfnamefont {I.}~\bibnamefont {Kerenidis}}, \ and\ \bibinfo
  {author} {\bibfnamefont {S.}~\bibnamefont {Wehner}},\ }\href@noop {}
  {\bibfield  {journal} {\bibinfo  {journal} {arXiv:1610.05238}\ } (\bibinfo
  {year} {2016})},\ \Eprint {http://arxiv.org/abs/1610.05238} {arXiv:1610.05238
  [cs.NI]} \BibitemShut {NoStop}%
\bibitem [{\citenamefont {Chakraborty}\ \emph {et~al.}(2019)\citenamefont
  {Chakraborty}, \citenamefont {Rozpedek}, \citenamefont {Dahlberg},\ and\
  \citenamefont {Wehner}}]{chakrabortydistributedrouting}%
  \BibitemOpen
  \bibfield  {author} {\bibinfo {author} {\bibfnamefont {K.}~\bibnamefont
  {Chakraborty}}, \bibinfo {author} {\bibfnamefont {F.}~\bibnamefont
  {Rozpedek}}, \bibinfo {author} {\bibfnamefont {A.}~\bibnamefont {Dahlberg}},
  \ and\ \bibinfo {author} {\bibfnamefont {S.}~\bibnamefont {Wehner}},\ }\href
  {https://arxiv.org/abs/1907.11630} {\bibfield  {journal} {\bibinfo  {journal}
  {arXiv:1907.11630}\ } (\bibinfo {year} {2019})}\BibitemShut {NoStop}%
\bibitem [{\citenamefont {Chakraborty}\ \emph {et~al.}(2020)\citenamefont
  {Chakraborty}, \citenamefont {Elkouss}, \citenamefont {Rijsman},\ and\
  \citenamefont {Wehner}}]{wehnermulticommodity}%
  \BibitemOpen
  \bibfield  {author} {\bibinfo {author} {\bibfnamefont {K.}~\bibnamefont
  {Chakraborty}}, \bibinfo {author} {\bibfnamefont {D.}~\bibnamefont
  {Elkouss}}, \bibinfo {author} {\bibfnamefont {B.}~\bibnamefont {Rijsman}}, \
  and\ \bibinfo {author} {\bibfnamefont {S.}~\bibnamefont {Wehner}},\ }\href
  {https://arxiv.org/abs/2005.14304} {\bibfield  {journal} {\bibinfo  {journal}
  {arXiv:2005.14304}\ } (\bibinfo {year} {2020})}\BibitemShut {NoStop}%
\bibitem [{\citenamefont {Van~Meter}\ \emph {et~al.}(2013)\citenamefont
  {Van~Meter}, \citenamefont {Satoh}, \citenamefont {Ladd}, \citenamefont
  {Munro},\ and\ \citenamefont {Nemoto}}]{vanmeterrouting}%
  \BibitemOpen
  \bibfield  {author} {\bibinfo {author} {\bibfnamefont {R.}~\bibnamefont
  {Van~Meter}}, \bibinfo {author} {\bibfnamefont {T.}~\bibnamefont {Satoh}},
  \bibinfo {author} {\bibfnamefont {T.~D.}\ \bibnamefont {Ladd}}, \bibinfo
  {author} {\bibfnamefont {W.~J.}\ \bibnamefont {Munro}}, \ and\ \bibinfo
  {author} {\bibfnamefont {K.}~\bibnamefont {Nemoto}},\ }\href
  {https://link.springer.com/article/10.1007} {\bibfield  {journal} {\bibinfo
  {journal} {Networking Science}\ }\textbf {\bibinfo {volume} {3}} (\bibinfo
  {year} {2013})}\BibitemShut {NoStop}%
\bibitem [{\citenamefont {Chang}\ \emph {et~al.}(2019)\citenamefont {Chang},
  \citenamefont {Li}, \citenamefont {Wu}, \citenamefont {Jiang}, \citenamefont
  {Zhang}, \citenamefont {Pu}, \citenamefont {Chang},\ and\ \citenamefont
  {Duan}}]{multiplex}%
  \BibitemOpen
  \bibfield  {author} {\bibinfo {author} {\bibfnamefont {W.}~\bibnamefont
  {Chang}}, \bibinfo {author} {\bibfnamefont {C.}~\bibnamefont {Li}}, \bibinfo
  {author} {\bibfnamefont {Y.-K.}\ \bibnamefont {Wu}}, \bibinfo {author}
  {\bibfnamefont {N.}~\bibnamefont {Jiang}}, \bibinfo {author} {\bibfnamefont
  {S.}~\bibnamefont {Zhang}}, \bibinfo {author} {\bibfnamefont {Y.-F.}\
  \bibnamefont {Pu}}, \bibinfo {author} {\bibfnamefont {X.-Y.}\ \bibnamefont
  {Chang}}, \ and\ \bibinfo {author} {\bibfnamefont {L.-M.}\ \bibnamefont
  {Duan}},\ }\href {\doibase 10.1103/PhysRevX.9.041033} {\bibfield  {journal}
  {\bibinfo  {journal} {Physical Review X}\ }\textbf {\bibinfo {volume} {9}},\
  \bibinfo {pages} {041033} (\bibinfo {year} {2019})}\BibitemShut {NoStop}%
\bibitem [{\citenamefont {Werner}(1989)}]{werner}%
  \BibitemOpen
  \bibfield  {author} {\bibinfo {author} {\bibfnamefont {R.~F.}\ \bibnamefont
  {Werner}},\ }\href {\doibase 10.1103/PhysRevA.40.4277} {\bibfield  {journal}
  {\bibinfo  {journal} {Physical Review A}\ }\textbf {\bibinfo {volume} {40}},\
  \bibinfo {pages} {4277} (\bibinfo {year} {1989})}\BibitemShut {NoStop}%
\bibitem [{\citenamefont {Ning}\ \emph {et~al.}(2019)\citenamefont {Ning},
  \citenamefont {Huang}, \citenamefont {Han}, \citenamefont {Li}, \citenamefont
  {Deng}, \citenamefont {Yang}, \citenamefont {Zhong}, \citenamefont {Xia},
  \citenamefont {Xu}, \citenamefont {Zheng},\ and\ \citenamefont
  {Zheng}}]{biao2019swap}%
  \BibitemOpen
  \bibfield  {author} {\bibinfo {author} {\bibfnamefont {W.}~\bibnamefont
  {Ning}}, \bibinfo {author} {\bibfnamefont {X.-J.}\ \bibnamefont {Huang}},
  \bibinfo {author} {\bibfnamefont {P.-R.}\ \bibnamefont {Han}}, \bibinfo
  {author} {\bibfnamefont {H.}~\bibnamefont {Li}}, \bibinfo {author}
  {\bibfnamefont {H.}~\bibnamefont {Deng}}, \bibinfo {author} {\bibfnamefont
  {Z.-B.}\ \bibnamefont {Yang}}, \bibinfo {author} {\bibfnamefont {Z.-R.}\
  \bibnamefont {Zhong}}, \bibinfo {author} {\bibfnamefont {Y.}~\bibnamefont
  {Xia}}, \bibinfo {author} {\bibfnamefont {K.}~\bibnamefont {Xu}}, \bibinfo
  {author} {\bibfnamefont {D.}~\bibnamefont {Zheng}}, \ and\ \bibinfo {author}
  {\bibfnamefont {S.-B.}\ \bibnamefont {Zheng}},\ }\href {\doibase
  10.1103/PhysRevLett.123.060502} {\bibfield  {journal} {\bibinfo  {journal}
  {Physical Review Letters}\ }\textbf {\bibinfo {volume} {123}},\ \bibinfo
  {pages} {060502} (\bibinfo {year} {2019})}\BibitemShut {NoStop}%
\bibitem [{\citenamefont {D{\"u}r}\ \emph {et~al.}(1999)\citenamefont
  {D{\"u}r}, \citenamefont {Briegel}, \citenamefont {Cirac},\ and\
  \citenamefont {Zoller}}]{dur1999quantum}%
  \BibitemOpen
  \bibfield  {author} {\bibinfo {author} {\bibfnamefont {W.}~\bibnamefont
  {D{\"u}r}}, \bibinfo {author} {\bibfnamefont {H.-J.}\ \bibnamefont
  {Briegel}}, \bibinfo {author} {\bibfnamefont {J.~I.}\ \bibnamefont {Cirac}},
  \ and\ \bibinfo {author} {\bibfnamefont {P.}~\bibnamefont {Zoller}},\ }\href
  {https://journals.aps.org/pra/pdf/10.1103/PhysRevA.59.169} {\bibfield
  {journal} {\bibinfo  {journal} {Physical Review A}\ }\textbf {\bibinfo
  {volume} {59}},\ \bibinfo {pages} {169} (\bibinfo {year} {1999})}\BibitemShut
  {NoStop}%
\bibitem [{\citenamefont {Muralidharan}\ \emph {et~al.}(2016)\citenamefont
  {Muralidharan}, \citenamefont {Li}, \citenamefont {Kim}, \citenamefont
  {L{\"u}tkenhaus}, \citenamefont {Lukin},\ and\ \citenamefont
  {Jiang}}]{muralidharan2016optimal}%
  \BibitemOpen
  \bibfield  {author} {\bibinfo {author} {\bibfnamefont {S.}~\bibnamefont
  {Muralidharan}}, \bibinfo {author} {\bibfnamefont {L.}~\bibnamefont {Li}},
  \bibinfo {author} {\bibfnamefont {J.}~\bibnamefont {Kim}}, \bibinfo {author}
  {\bibfnamefont {N.}~\bibnamefont {L{\"u}tkenhaus}}, \bibinfo {author}
  {\bibfnamefont {M.~D.}\ \bibnamefont {Lukin}}, \ and\ \bibinfo {author}
  {\bibfnamefont {L.}~\bibnamefont {Jiang}},\ }\href
  {https://www.nature.com/articles/srep20463} {\bibfield  {journal} {\bibinfo
  {journal} {Scientific reports}\ }\textbf {\bibinfo {volume} {6}},\ \bibinfo
  {pages} {20463} (\bibinfo {year} {2016})}\BibitemShut {NoStop}%
\bibitem [{\citenamefont {Bennett}\ \emph
  {et~al.}(1996{\natexlab{c}})\citenamefont {Bennett}, \citenamefont
  {DiVincenzo}, \citenamefont {Smolin},\ and\ \citenamefont
  {Wootters}}]{hashingrate}%
  \BibitemOpen
  \bibfield  {author} {\bibinfo {author} {\bibfnamefont {C.~H.}\ \bibnamefont
  {Bennett}}, \bibinfo {author} {\bibfnamefont {D.~P.}\ \bibnamefont
  {DiVincenzo}}, \bibinfo {author} {\bibfnamefont {J.~A.}\ \bibnamefont
  {Smolin}}, \ and\ \bibinfo {author} {\bibfnamefont {W.~K.}\ \bibnamefont
  {Wootters}},\ }\href {\doibase 10.1103/PhysRevA.54.3824} {\bibfield
  {journal} {\bibinfo  {journal} {Physical Review A}\ }\textbf {\bibinfo
  {volume} {54}},\ \bibinfo {pages} {3824} (\bibinfo {year}
  {1996}{\natexlab{c}})}\BibitemShut {NoStop}%
\bibitem [{\citenamefont {Murta}\ \emph {et~al.}(2020)\citenamefont {Murta},
  \citenamefont {Rozp\ifmmode~\mbox{\k{e}}\else \k{e}\fi{}dek}, \citenamefont
  {Ribeiro}, \citenamefont {Elkouss},\ and\ \citenamefont
  {Wehner}}]{wehner2020keyrates}%
  \BibitemOpen
  \bibfield  {author} {\bibinfo {author} {\bibfnamefont {G.}~\bibnamefont
  {Murta}}, \bibinfo {author} {\bibfnamefont {F.}~\bibnamefont
  {Rozp\ifmmode~\mbox{\k{e}}\else \k{e}\fi{}dek}}, \bibinfo {author}
  {\bibfnamefont {J.}~\bibnamefont {Ribeiro}}, \bibinfo {author} {\bibfnamefont
  {D.}~\bibnamefont {Elkouss}}, \ and\ \bibinfo {author} {\bibfnamefont
  {S.}~\bibnamefont {Wehner}},\ }\href {\doibase 10.1103/PhysRevA.101.062321}
  {\bibfield  {journal} {\bibinfo  {journal} {Physical Review A}\ }\textbf
  {\bibinfo {volume} {101}},\ \bibinfo {pages} {062321} (\bibinfo {year}
  {2020})}\BibitemShut {NoStop}%
\bibitem [{\citenamefont {D{\"u}r}\ and\ \citenamefont
  {Briegel}(2007)}]{dur2007entanglement}%
  \BibitemOpen
  \bibfield  {author} {\bibinfo {author} {\bibfnamefont {W.}~\bibnamefont
  {D{\"u}r}}\ and\ \bibinfo {author} {\bibfnamefont {H.~J.}\ \bibnamefont
  {Briegel}},\ }\href
  {https://iopscience.iop.org/article/10.1088/0034-4885/70/8/R03/meta}
  {\bibfield  {journal} {\bibinfo  {journal} {Reports on Progress in Physics}\
  }\textbf {\bibinfo {volume} {70}},\ \bibinfo {pages} {1381} (\bibinfo {year}
  {2007})}\BibitemShut {NoStop}%
\bibitem [{\citenamefont {Xu}\ \emph {et~al.}(2019)\citenamefont {Xu},
  \citenamefont {Tao}, \citenamefont {Wu}, \citenamefont {Ye},\ and\
  \citenamefont {Zhang}}]{zhang2019constrainedrouting}%
  \BibitemOpen
  \bibfield  {author} {\bibinfo {author} {\bibfnamefont {C.}~\bibnamefont
  {Xu}}, \bibinfo {author} {\bibfnamefont {L.}~\bibnamefont {Tao}}, \bibinfo
  {author} {\bibfnamefont {H.}~\bibnamefont {Wu}}, \bibinfo {author}
  {\bibfnamefont {D.}~\bibnamefont {Ye}}, \ and\ \bibinfo {author}
  {\bibfnamefont {G.}~\bibnamefont {Zhang}},\ }\href
  {https://arxiv.org/pdf/1902.10312.pdf} {\bibfield  {journal} {\bibinfo
  {journal} {arXiv}\ }\textbf {\bibinfo {volume} {arXiv:1902.10312}} (\bibinfo
  {year} {2019})}\BibitemShut {NoStop}%
\bibitem [{\citenamefont {Das}\ \emph {et~al.}(2018)\citenamefont {Das},
  \citenamefont {Khatri},\ and\ \citenamefont
  {Dowling}}]{dowling2018networktopology}%
  \BibitemOpen
  \bibfield  {author} {\bibinfo {author} {\bibfnamefont {S.}~\bibnamefont
  {Das}}, \bibinfo {author} {\bibfnamefont {S.}~\bibnamefont {Khatri}}, \ and\
  \bibinfo {author} {\bibfnamefont {J.~P.}\ \bibnamefont {Dowling}},\ }\href
  {\doibase 10.1103/PhysRevA.97.012335} {\bibfield  {journal} {\bibinfo
  {journal} {Phys. Rev. A}\ }\textbf {\bibinfo {volume} {97}},\ \bibinfo
  {pages} {012335} (\bibinfo {year} {2018})}\BibitemShut {NoStop}%
\bibitem [{\citenamefont {\AA{}berg}\ \emph {et~al.}(2020)\citenamefont
  {\AA{}berg}, \citenamefont {Nery}, \citenamefont {Duarte},\ and\
  \citenamefont {Chaves}}]{chaves2020networktopology}%
  \BibitemOpen
  \bibfield  {author} {\bibinfo {author} {\bibfnamefont {J.}~\bibnamefont
  {\AA{}berg}}, \bibinfo {author} {\bibfnamefont {R.}~\bibnamefont {Nery}},
  \bibinfo {author} {\bibfnamefont {C.}~\bibnamefont {Duarte}}, \ and\ \bibinfo
  {author} {\bibfnamefont {R.}~\bibnamefont {Chaves}},\ }\href {\doibase
  10.1103/PhysRevLett.125.110505} {\bibfield  {journal} {\bibinfo  {journal}
  {Phys. Rev. Lett.}\ }\textbf {\bibinfo {volume} {125}},\ \bibinfo {pages}
  {110505} (\bibinfo {year} {2020})}\BibitemShut {NoStop}%
\bibitem [{\citenamefont {Sykes}(1964)}]{sykes1964percolation}%
  \BibitemOpen
  \bibfield  {author} {\bibinfo {author} {\bibfnamefont {M.~F.}\ \bibnamefont
  {Sykes}},\ }\href {\doibase 10.1063/1.1704215} {\bibfield  {journal}
  {\bibinfo  {journal} {Journal of Material Physics}\ }\textbf {\bibinfo
  {volume} {5}},\ \bibinfo {pages} {1117} (\bibinfo {year} {1964})}\BibitemShut
  {NoStop}%
\bibitem [{\citenamefont {Hagberg}\ \emph {et~al.}(2008)\citenamefont
  {Hagberg}, \citenamefont {Schult},\ and\ \citenamefont
  {Swart}}]{swart2008networkx}%
  \BibitemOpen
  \bibfield  {author} {\bibinfo {author} {\bibfnamefont {A.~A.}\ \bibnamefont
  {Hagberg}}, \bibinfo {author} {\bibfnamefont {D.~A.}\ \bibnamefont {Schult}},
  \ and\ \bibinfo {author} {\bibfnamefont {P.~J.}\ \bibnamefont {Swart}},\ }in\
  \href@noop {} {\emph {\bibinfo {booktitle} {Proceedings of the 7th Python in
  Science Conference (SciPy 2008)}}},\ \bibinfo {editor} {edited by\ \bibinfo
  {editor} {\bibfnamefont {G.}~\bibnamefont {Varoquaux}}, \bibinfo {editor}
  {\bibfnamefont {T.}~\bibnamefont {Vaught}}, \ and\ \bibinfo {editor}
  {\bibfnamefont {J.}~\bibnamefont {Millman}}}\ (\bibinfo {year} {2008})\ pp.\
  \bibinfo {pages} {11--15}\BibitemShut {NoStop}%
\end{thebibliography}%

\end{document}